%% file: main.tex
\documentclass[11pt]{article}
\usepackage{graphicx} 
\usepackage{caption}
\usepackage{subcaption}
\usepackage{stfloats}
\usepackage{amsmath,amsthm,amssymb,amsfonts,mathtools,bbold}
\usepackage{hyperref}
\usepackage{xcolor}
\usepackage{enumerate}
\usepackage{cite}
\usepackage{tikz}
\usepackage{color}
\usetikzlibrary{positioning}
\usetikzlibrary{decorations.pathreplacing}

\newcommand{\cO}{\mathcal{O}}
\newcommand{\cI}{\mathcal{I}}

\newcommand{\hs}{\mathcal{H}}

\definecolor{brown}{HTML}{8B4513}
\definecolor{blue}{HTML}{007Aff}
\definecolor{green}{HTML}{94E700}
\definecolor{orange}{HTML}{FF8800}
\definecolor{red}{HTML}{FF0000}
\definecolor{pink}{HTML}{FF9797}
\definecolor{darkblue}{HTML}{03468F}
\definecolor{darkgreen}{HTML}{007355}
\definecolor{purple}{HTML}{82218b}

\begin{document}

\begin{titlepage}
{\ }
\vskip 1in

\begin{center}
{\LARGE{Tensor networks for black hole interiors: \\[0.3cm]
non-isometries, quantum extremal surfaces, and wormholes}}
\vskip 0.5in {\Large Gracemarie Bueller,\footnote{gracemarie.bueller@colorado.edu} Oliver DeWolfe,\footnote{oliver.dewolfe@colorado.edu} and Kenneth Higginbotham\footnote{kenneth.higginbotham@colorado.edu}}
\vskip 0.2in {\it Department of Physics and \\
Center for Theory of Quantum Matter \\
390 UCB \\ University of Colorado \\ Boulder, CO 80309, USA}
\end{center}
\vskip 0.5in

\begin{abstract}\noindent
We use hyperbolic tensor networks to construct a holographic map for black hole interiors that adds a notion of locality to the non-isometric codes proposed by Akers, Engelhardt, Harlow, Penington, and Vardhan. We use tools provided by these networks to study the relationship between non-isometries and quantum extremal surfaces behind the horizon. Furthermore, we introduce a limited notion of dynamics for these interior tensor networks based on the qudit models introduced by Akers et al., and study the evolution of quantum extremal surfaces in an evaporating black hole. We also find a tensor network description of a wormhole connecting the black hole interior to the radiation, providing a mechanism for interior states and operators to be encoded in the radiation after the Page time. As a particular case, we construct a tensor network realization of the backwards-forwards maps recently proposed to incorporate non-trivial effective dynamics in dynamical constructions of these non-isometric black hole codes.
\end{abstract}

\end{titlepage}

\section{Introduction}

The work of Akers, Engelhardt, Harlow, Penington, and Vardhan in \cite{akers_black_2022} (which we will refer to non-alphabetically but euphoniously as PHEVA) demonstrated that non-isometric codes can be used to reproduce a significant array of results from the field of black hole information. Most notably, they described non-isometric holographic maps for simple qudit models of black holes capable of reproducing the quantum extremal surface (QES) formula \cite{ryu_holographic_2006,faulkner_quantum_2013,engelhardt_quantum_2015,penington_entanglement_2020,almheiri_entropy_2019} and obtaining the Page curve \cite{page_information_1993}.  This represented a significant step forward in our understanding of how semi-classical gravity may arise from a quantum theory of gravity, and it is our hope that we can improve this understanding by improving on these black hole qudit models.

This work takes one step towards a more realistic qudit model of the black hole interior. The model used in \cite{akers_black_2022} for the semi-classical interior implemented a limited, binary notion of locality: you're either behind the black hole horizon or you're not. We aim to include a slightly more refined notion of geometry by allowing for relative positions of degrees of freedom in the interior and exterior of the black hole. Holographic tensor networks are a natural candidate -- using qudits, they describe bulk locality built from the quantum informational properties of the boundary. The geometry provided by such tensor network constructions makes available tools such as Ryu-Takayanagi (RT)-like formulas for studying QESs and operator reconstruction within entanglement wedges.

Tensor network descriptions of AdS/CFT were pioneered by \cite{swingle_entanglement_2012}, where a discrete bulk was built from $1d$ critical Ising models using MERA tensor networks. Later, the HaPPy code \cite{pastawski_holographic_2015} was created to incorporate the error correcting properties of AdS/CFT \cite{almheiri_bulk_2015} into a holographic tensor network. Built out of perfect tensors, the HaPPy code was designed to act as an exact isometry on bulk input states, encoding them redundantly in the boundary. These constructions focused on regions exterior to any horizon and represented black holes by excising tensors in their interior, leaving dangling legs representing \textit{fundamental} black hole degrees of freedom along the horizon. Viewing these legs as inputs, the remaining annular-shaped tensor network would isometrically encode the fundamental black hole degrees of freedom on the horizon in the CFT degrees of freedom on the boundary. In this work, we will refer to such tensor network codes outside the horizon as the ``outer code.''

However, tensor networks need not only be useful for the exterior of the black hole. An observer falling past the horizon should still be able to give a local description of the surroundings, even when they've passed into the interior. Thus the locality provided by a tensor network should also be at least approximately applicable to the interior of a black hole. A great deal of work has focused on the application of tensor networks to black hole interiors. For example, tensor networks have been used to study the interior of two-sided black holes \cite{hartman_time_2013}, the complexity of black hole interiors \cite{brown_complexity_2016}, and in constructions of the python's lunch proposal \cite{brown_pythons_2020}. Furthermore, random tensor networks constructed in \cite{hayden_holographic_2016} were used to observe phase transitions in minimal surfaces consistent with the formation of a black hole. All of these examples suggest that the excised portions of a holographic tensor network behind a horizon can be filled in to give a local description of black hole interiors.

There is also strong evidence that tensor networks describing black holes must involve non-isometries in a similar way to PHEVA's holographic codes. While minimal surfaces in the random tensor networks of \cite{hayden_holographic_2016} underwent a black hole-like phase transition, the tensor network transitioned from a bulk-to-boundary isometry to a boundary-to-bulk isometry. Therefore, the appearance of black hole features in the random tensor network seems to correspond to a bulk-to-boundary map that must be non-isometric. More recently, OPE data from 2D CFTs was used to demonstrate the existence of non-isometries in holographic random tensor networks \cite{chandra_toward_2023}.

We therefore construct a tensor network description for black hole interiors to give a notion of locality to PHEVA's non-isometric code. This interior tensor network will define an ``inner code'' mapping bulk \textit{effective} degrees of freedom in the black hole interior to \textit{fundamental} black hole degrees of freedom on the horizon. The outer code then isometrically encodes the fundamental black hole degrees of freedom in the CFT degrees of freedom on the boundary. The entire holographic code is then given by the concatenation of the inner and outer codes. Because outer codes are already well understood, we will focus on the inner code as the generalization of PHEVA's work. 

Non-isometry will be incorporated in the inner code by allowing for an arbitrary number of bulk inputs into each tensor, requiring both isometric and non-isometric tensors. The more refined notion of locality will allow us to study the geometric structure of non-isometries in the map. With the added ability to study QESs through the generalized entropy of inner code states, we will make a new connection between non-isometries and QESs in the inner code.

Furthermore, we will use the dynamics of effective and fundamental states defined in \cite{akers_black_2022} and refined in \cite{dewolfe_non-isometric_2023} to define a limited notion of time evolution for these interior tensor networks. This will allow us to study the phase transition between two candidate QESs in an evaporating black hole. We will also show that ``dynamically generated states'' (states prepared by the effective dynamics from non-singular initial data) can be used to construct tensor networks that exhibit wormholes connecting the interior to a reservoir where the radiation is stored. This provides a mechanism that could be used to reconstruct interior operators in the radiation instead of on the boundary of the AdS bulk long after the Page time, as expected from the work of \cite{penington_entanglement_2020,almheiri_entropy_2019}. 

Finally, we will use our model of dynamics to construct a version of the backwards-forwards holographic maps \cite{dewolfe_non-isometric_2023,dewolfe_bulk_2024} that has the same notion of locality. The backwards-forwards (BF) and backwards-forwards-post-selection (BFP) maps were initially proposed in \cite{dewolfe_non-isometric_2023} as a generalization of PHEVA's dynamical holographic maps (also introduced in \cite{akers_black_2022}) to include non-trivial dynamics behind the horizon while avoiding issues with unitarity and computational complexity brought up by Kim and Preskill in \cite{kim_complementarity_2023}. Just as PHEVA viewed their dynamical map as a special example of their more general non-isometric code, we view these local constructions of the BF and BFP maps presented here as special examples of the more general interior tensor network inner codes discussed earlier.

The remainder of this work will be organized as follows. Sec.~\ref{sec:tessellations} will use hyperbolic tessellations to assign a notion of locality (on an AdS length scale) to bulk degrees of freedom. Sec.~\ref{sec:TN} will construct a tensor network description of the interior from this tessellated description of the bulk. We will study the structure of non-isometries, QESs, and their connection here. Sec.~\ref{sec:dynamics} will introduce a limited notion of dynamics in order to study the evolution of QESs for an evaporating black hole and provide comments on interior operator reconstruction. Sec.~\ref{sec:BFP} will give the constructions of the BF and BFP maps from the dynamics defined in Sec.~\ref{sec:dynamics}. Finally, we will provide closing remarks and future directions in Sec.~\ref{sec:conc}.

\section{Locality from hyperbolic tessellations} \label{sec:tessellations}

\subsection{Review of qudit black hole models}

We begin with a brief review of the black hole model established by PHEVA in \cite{akers_black_2022}. Two descriptions of the black hole are given, both using $D$-level qudits in finite dimensional Hilbert spaces. The \textit{effective description} gives the perspective of an observer falling into a black hole in a semi-classical bulk. Interior degrees of freedom behind the horizon are divided into left-moving (radially ingoing) modes $\ell$ and right-moving (radially outgoing) modes $r$. The left-movers $\ell$ keep track of what fell into the black hole from the exterior, while the right-movers $r$ represent interior Hawking pairs that are maximally entangled with outgoing radiation.

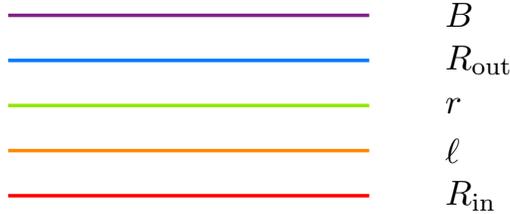
\begin{figure}
    \centering
    \input{colors}
    \caption{The color scheme used to identify qudits in all diagrams throughout this work.}
    \label{fig:colors}
\end{figure}

The \textit{fundamental description} provides the account of an exterior (or boundary) observer watching the infalling observer from a safe distance. They describe the black hole using fundamental quantum gravity degrees of freedom $B$; the number of $B$ qudits is determined by the area of the event horizon. Both descriptions share the same exterior since the descriptions of both observers should agree outside of the horizon. Any exterior degrees of freedom are removed to a reservoir $R$, which is divided into two parts: infalling modes $R_\text{in}$ that will eventually cross the horizon to become $\ell$ modes in the effective description, and outgoing Hawking pairs $R_\text{out}$ that are maximally entangled with the $r$ modes in the effective description. The resulting Hilbert spaces for both descriptions are,
\begin{equation}
    \hs_\text{eff} = \hs_\ell \otimes \hs_r \otimes \hs_{R_\text{in}} \otimes \hs_{R_\text{out}}, \qquad \hs_\text{fun} = \hs_B \otimes \hs_{R_\text{in}} \otimes \hs_{R_\text{out}},
\end{equation}
where each factor is comprised of $D$-dimensional qudits. Throughout this paper we will use colors to visually distinguish between these various Hilbert space factors in diagrams, depicted in Fig.~\ref{fig:colors}.

The goal of PHEVA's work in \cite{akers_black_2022} was to create a holographic map $V$ from the interior degrees of freedom in the effective description to the fundamental black hole degrees of freedom:
\begin{equation} \label{eq:holo_map}
    V \,:\, \ell r \rightarrow B
\end{equation}
For a black hole that is evaporating, the size of $\hs_B$ is always decreasing as more Hawking radiation leaves into $\hs_{R_\text{out}}$ than enters from $\hs_{R_\text{in}}$, and the area of the horizon shrinks. From the point of view of the effective description however, the size of $\hs_\ell\otimes\hs_r$ is always increasing as more Hawking pairs $rR_\text{out}$ are generated.  Therefore, there will be a time (close to the Page time) when there are more effective interior degrees of freedom $\ell r$ than fundamental black hole degrees of freedom $B$, and $V$ must become non-isometric. Remarkably, \cite{akers_black_2022} demonstrated that this non-isometry is not something to be feared, but is instead necessary to reproduce many key features of black hole information physics.

As a model of the semi-classical bulk description of a black hole, geometry should be fully incorporated into the effective description. (The same is not true of the fundamental description, where geometry is assumed to be emergent.) As the non-isometric codes initiated by PHEVA currently stand, these qudit models do not incorporate geometry. There is only the binary distinction of ``inside the black hole'' ($\ell r$) or ``outside the black hole'' ($R_\text{in}R_\text{out}$) and direction of propagation inward ($\ell R_\text{in}$) vs.~outward ($r R_\text{out}$). No further notion of locality exists.

The primary goal of this work is to take the first step in adding a notion of locality to the non-isometric holographic codes proposed by \cite{akers_black_2022}. To do so, we will specialize to AdS/CFT as a specific example of a holographic duality\footnote{Because of this, we will often use the words ``bulk''/``effective'' and ``boundary''/``fundamental'' interchangeably.} and make use of hyperbolic tessellations. As a concrete case, we will consider tessellations of AdS$_3$ spatial slices, but we expect the arguments should be generalizable to higher dimensions. 

\subsection{Hyperbolic tessellations and degrees of freedom}

\begin{figure}
    \centering
    \includegraphics[width=0.5\linewidth]{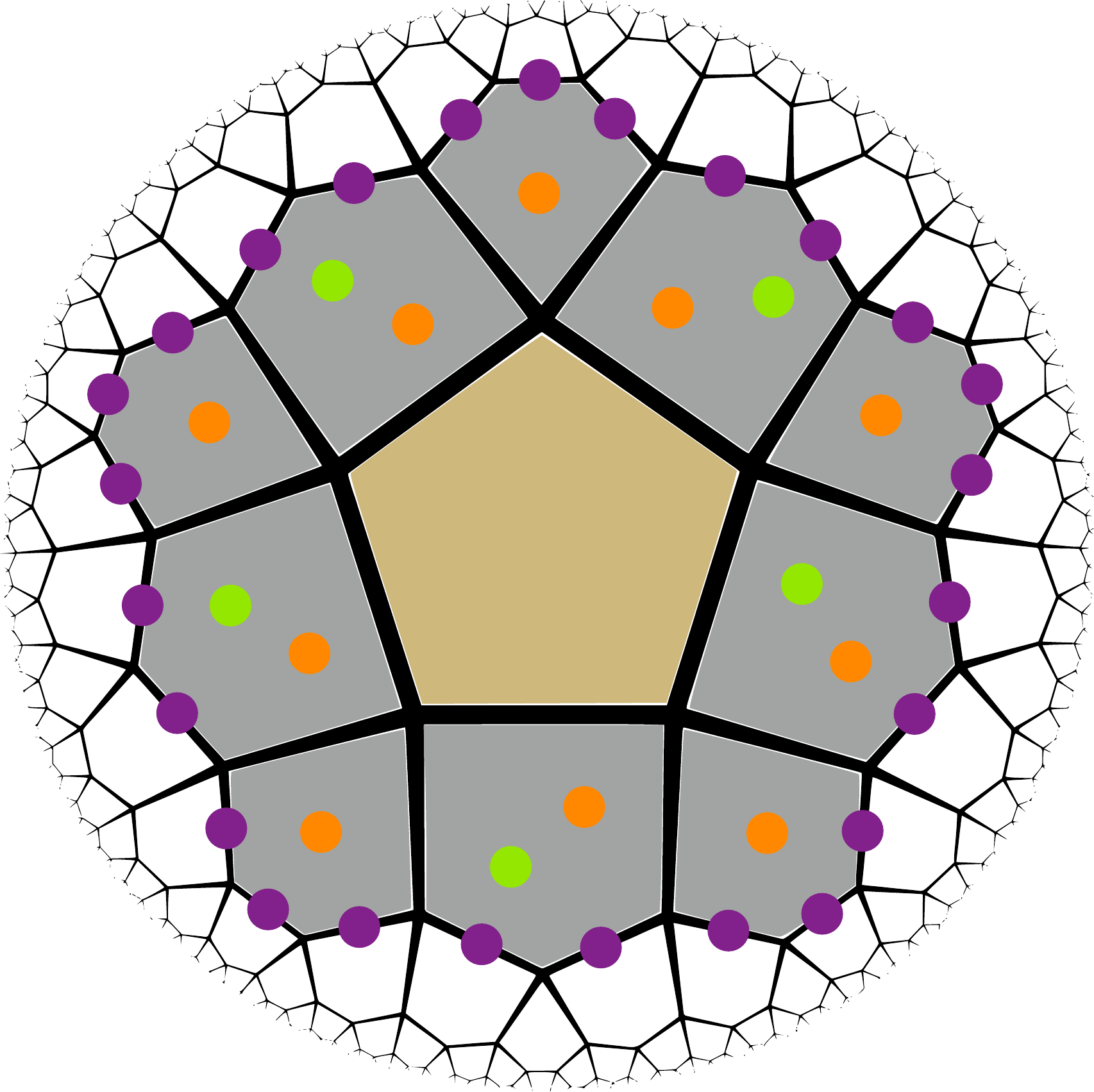}
    \caption{A $\{5,4\}$ hyperbolic tessellation created by placing four pentagons around each vertex. Shaded cells describe the black hole interior and are colored according to their row: row 0 is colored gold and row 1 is colored gray. Qudits are represented as dots and colored by type according to Fig.~\ref{fig:colors}. $r$ (\textcolor{green}{green}) and $\ell$ (\textcolor{orange}{orange}) qudits are associated to cells, while $B$ (\textcolor{purple}{purple}) qudits are associated to edges along the horizon $\Lambda_\text{BH}$.}
    \label{fig:vertex_inflation}
\end{figure}

Regular tessellations are tilings of a space using copies of a single regular polygon. 2-dimensional regular tessellations can be defined by the Schl\"{a}fli symbol $\{n,k\}$, where $n$ is the number of sides of the defining polygon and $k$ is the number of polygons around each vertex. Fig.~\ref{fig:vertex_inflation} depicts a $\{5,4\}$ hyperbolic tessellation: each cell is a pentagon ($n=5$) and there are four around each vertex ($k=4$). 

By comparing the sum of the interior angles of the $k$ polygons around a vertex to $2\pi$, a notion of scalar curvature can be defined for a tessellation:
\begin{equation} \label{eq:curvature}
    R_\text{tess} = 1 - \frac{k(n-2)}{2n}.
\end{equation}
Positive curvature is indicated by $R_\text{tess}>0$, negative by $R_\text{tess}<0$, and flat tessellations are given by $R_\text{tess}=0$. The tessellation of a spacetime metric can then be constructed by comparing this curvature with the metric's Ricci scalar curvature. We note that since the AdS-Schwarzschild metric is a vacuum solution of Einstein's equations with a cosmological constant, it has the same Ricci scalar curvature as the regular AdS metric,
\begin{equation}
    R = -\frac{d(d-1)}{L^2}, \quad \text{for AdS$_d$},
\end{equation}
with $L$ the AdS length scale. Therefore, spatial slices of both AdS and AdS-Schwarzschild metrics have the same tessellation. Both require $k < 2n/(n-2)$ to be negatively curved; such tessellations are referred to as ``hyperbolic'' tessellations.

We will find it useful to parameterize cells of the interior tessellation according to rows given by ``vertex inflation'' \cite{jahn_holographic_2021} -- a central cell defines row 0, and row $N$ consists of all cells not in any earlier row that share a corner with row $N-1$. The cells in the pentagonal tessellation of Fig.~\ref{fig:vertex_inflation} are colored according to their row: row 0 is colored gold and row 1 is colored gray. This permits a method for cell and edge counting by row, described in Appendix~\ref{app:counting}. For the majority of tessellations, each cell in a row connects to the row beneath it by only an edge or a vertex.\footnote{For tessellations with $k=3$, cells will be connected to the row beneath it by either one or two edges, not one vertex. In this case, $f_v$ in (\ref{eq:facevec}) can be negative. Despite this, the formulas for the total number of cells and number of edges on a row give the correct results.} Denoting the number of edge connected cells as $f_e$ and the number of vertex connected cells as $f_v$, we may count the number of cells in row $N$ of an $\{n,k\}$ tessellation by
\begin{equation} \label{eq:facevec}
    \begin{pmatrix}
        f_e \\
        f_v
    \end{pmatrix} = 
    \begin{pmatrix}
        n-3                 &   n-2 \\
        (n-4)(k-3) + (k-4)  &   (n-3)(k-3) + (k-4) 
    \end{pmatrix}^{N-1}
    \begin{pmatrix}
        n \\
        n(k-3)
    \end{pmatrix}, \quad N>0.
\end{equation}
The total number of cells in row $N$ is then given by
\begin{equation}
    \#\text{ cells in row } N =
    \begin{pmatrix}
        1   &   1
    \end{pmatrix}
    \begin{pmatrix}
        f_e \\
        f_v
    \end{pmatrix},
\end{equation}
while the total number of edges on the outside of row $N$ is given by
\begin{equation} \label{eq:num_edges}
    \#\text{ edges on row } N = 
    \begin{pmatrix}
        n-3 &   n-2
    \end{pmatrix}
    \begin{pmatrix}
        f_e \\
        f_v
    \end{pmatrix}.
\end{equation}
We note that in the case of flat tessellations when  $k= 2n/(n-2)$, (\ref{eq:num_edges}) simplifies considerably to $n(1+2N)$.

We can now use these hyperbolic tessellations to give a notion of locality to the effective and fundamental degrees of freedom described above. Qudits describing bulk effective degrees of freedom will be associated to cells of the tessellation. To be more specific, we must identify the portions of the tessellation that are in the interior of the black hole. Because rows are roughly spherically symmetric, we will define a black hole interior by the number of rows it encompasses. For example, a black hole encompassing two rows ($N=0,1$) of a $\{5,4\}$ tessellation would contain all shaded cells in Fig.~\ref{fig:vertex_inflation}. We will denote the boundary between the interior cells and the exterior cells (defined by the edges on the outermost row of the black hole interior) as $\Lambda_\text{BH}$.\footnote{There is a freedom in where we choose $\Lambda_\text{BH}$, so long as it lives outside of the horizon \cite{penington_entanglement_2020}. Throughout this work, we will use this freedom to place $\Lambda_\text{BH}$ on the horizon; this is primarily a choice of convenience, as it will allow us to ignore unitary interactions between infalling modes $R_\text{in}$ and outgoing radiation $R_\text{out}$ that would be considered a part of the black hole evolution had we chosen $\Lambda_\text{BH}$ to lie away from the horizon.} 

Therefore, any bulk qudits within the cells behind $\Lambda_\text{BH}$ are considered to be in the interior of the black hole and categorized as either $\ell$ or $r$; any qudits outside are taken to be external to the black hole and categorized as $R_\text{in}$ or $R_\text{out}$. Following the work of \cite{almheiri_entropy_2019,penington_entanglement_2020}, we will allow the black hole to evaporate by removing $R_\text{in}$ and $R_\text{out}$ from the bulk AdS tessellation to an external reservoir $R$; this also allows us to restrict our attention to the portion of the tessellation in the interior of the black hole. The tessellation gives us a notion of locality to this effective description of the interior -- qudits in a given cell are all ``near'' each other, and the qudits in two adjacent cells are separated by a distance of $O(L)$. We note that this construction says nothing about the arrangement of qudits within a cell, since these hyperbolic tessellations do not describe sub-AdS length scale locality. \textcolor{green}{Green} and \textcolor{orange}{orange} dots in Fig.~\ref{fig:vertex_inflation} depict an example of $r$ and $\ell$ degrees of freedom (respectively) that are associated to interior cells of the $\{5,4\}$ tessellation. The placement of these qudits in Fig.~\ref{fig:vertex_inflation} is arbitrary but for simplicity chosen to respect the $2\pi/n$ rotational symmetry of the tessellation, which is a remnant of the spherical symmetry emerging in the limit that the black hole is much larger than the AdS length scale $L$.

Continuing our parametrization of the tessellation using rows, the total number of $\ell$ and $r$ qudits in a given row $N$ will be denoted as $q_\ell(N)$ and $q_r(N)$, respectively. For example, the state shown in Fig.~\ref{fig:vertex_inflation} has $q_\ell(0) = q_r(0) = 0$, $q_\ell(1) = 10$ and $q_r(1) = 5$. The size of each Hilbert space factor $\hs_i$ is then related to the sum of $q_i(N)$ over all rows,
\begin{equation}
    \sum_N q_\ell(N) = \log_D |\ell|, \qquad \sum_N q_r(N) = \log_D |r|,
\end{equation}
where we have used the shorthand $|\ell| \equiv |\hs_\ell|$ and $|r| \equiv |\hs_r|$ for the dimensions of the Hilbert space factors. We will use this shorthand for all Hilbert space factors in the remainder of this work.

The fundamental description of the black hole lives on the horizon, so $B$ will be given by placing one qudit on each edge of the tessellation along $\Lambda_\text{BH}$. The size of $B$ is then determined by the number of edges (\ref{eq:num_edges}) on the outermost row of the tessellation behind the interior,
\begin{equation} \label{eq:logB}
    q_B(N) = 
    \begin{pmatrix}
        n-3 &   n-2
    \end{pmatrix}
    \begin{pmatrix}
        f_e \\
        f_v
    \end{pmatrix},
\end{equation}
corresponding to the area of the horizon. The size of $\hs_B$ is then $\log_D |B| = q_B(N)$. For the example depicted in Fig.~\ref{fig:vertex_inflation}, $q_B(1)=25.$

This model constitutes a first step towards adding locality to the effective degrees of freedom $\ell r$ in the semi-classical black hole interior of PHEVA's model. We now wish to build a holographic map similar to (\ref{eq:holo_map}) of \cite{akers_black_2022} that will encode $\ell r$ in $B$ while incorporating this notion of locality. Tensor networks are a natural choice, and we now turn to the construction of a tensor network inner code for the black hole interior.

\section{Interior tensor network} \label{sec:TN}

Given a hyperbolic tessellation restricted to a black hole interior, a set of bulk degrees of freedom $\ell r$ associated to the cells of the tessellation, and fundamental degrees of freedom $B$ along the edges of $\Lambda_\text{BH}$, we construct an inner holographic code as a tensor network that maps $\ell r \rightarrow B$. The network will live on the lattice dual to the tessellation: for each cell there is a corresponding vertex $x$ in the dual lattice, while dual links connecting vertices are orthogonal to the edges of the tessellation. In this way, all but the outermost vertices will have $n$ links connecting them to neighboring vertices. Those on the edge of the interior tessellation will have uncontracted legs dangling across $\Lambda_\text{BH}$ -- these represent the fundamental $B$ degrees of freedom. Bulk uncontracted legs are then added to every vertex $x$ for each effective qudit $\ell r$ associated to the corresponding cell. 

Finally, tensors $T_x$ are assigned to each dual vertex $x$. The links of the dual lattice denote contractions between these tensors. The bulk uncontracted legs representing effective degrees of freedom will serve as inputs to each tensor, while the uncontracted boundary legs will serve as the outputs of the network. The directionality of contracted legs will depend on the flow of the network; we will always take the network to flow radially outward, with each row acting as successive layers of the map. Therefore, contracted legs on the outside of row $N$ will be considered outputs of the tensors in row $N$ and inputs to the tensors in row $N+1$. The direction of contracted legs between neighboring tensors within the same row can be ambiguous in some cases, but the choice of direction should not impact our results. For simplicity, we will assume that all legs, contracted and uncontracted, will have the same dimension $D$.

This gives a tensor network construction for the black hole interior, which we interpret as the ``inner code'' of the holographic map. A tensor network for the exterior of the black hole can be constructed in the usual way \cite{pastawski_holographic_2015,hayden_holographic_2016}, giving the ``outer code'' that completes the holographic map. Together, both codes form the holographic map $\ell r \rightarrow B \rightarrow \text{CFT}$. We take this outer code to act as an exact isometry, preserving all information in the $B$ qudits as they are encoded in the CFT degrees of freedom. We will therefore neglect the outer code in the remainder of this paper and focus our attention on the tensor network behind the horizon.

Thus focusing our attention, we note that we have made no restrictions on the number of effective degrees of freedom $\ell r$ used as inputs to each tensor $T_x$. This allows for the possibility of non-isometric tensors in the network that map a large number of uncontracted bulk legs to a small number of contracted legs. To model both isometric and non-isometric tensors, recall the general fact that any linear map can be written as an isometry followed by post-selection \cite{akers_black_2022}. Since every tensor $T_x$ in the network is a linear map from incoming legs to outgoing legs, they too can be put in this form. To be specific, let us define
\begin{align}
    \#_{\text{in},x} &\equiv \text{number legs ingoing to } T_x \\
    \#_{\text{out},x} &\equiv \text{number legs outgoing from } T_x.
\end{align}
For a tensor network built from an $\{n,k\}$ tessellation, the sum $\#_{\text{in},x} + \#_{\text{out},x}$ must be at least $n$ for every $x$.
If $\#_{\text{in},x} < \#_{\text{out},x}$, we take $T_x$ to be an exact isometry, constructed by inserting ancilla $A$ and performing some unitary $U_x$:
\begin{equation} \label{eq:T_iso}
    T_x = U_x |0\rangle_A, \quad \#_{\text{in},x} < \#_{\text{out},x}
\end{equation}
where the $A$ ancilla must have dimension
\begin{equation}
    \log_D |A| = \#_{\text{out},x} - \#_{\text{in},x}
\end{equation}
and $U_x$ has dimension $\#_{\text{out},x}$. Alternatively, tensors satisfying $\#_{\text{in},x} > \#_{\text{out},x}$ will be exact non-isometries with no inserted ancilla. We construct such non-isometric $T_x$ by performing a unitary $U_x$ on all incoming legs and then post-selecting:
\begin{equation} \label{eq:T_non-iso}
    T_x = \sqrt{|P|} \langle0|_P U_x, \quad \#_{\text{in},x} > \#_{\text{out},x},
\end{equation}
where the prefactor $\sqrt{|P|}$ has been included to preserve normalization. Here $\langle0|_P$ represents post-selection on a subsystem of dimension
\begin{equation}
    \log_D |P| = \#_{\text{in},x} - \#_{\text{out},x},
\end{equation}
while $U_x$ has dimension $\#_{\text{in},x}$.

The entire network will be built out of tensors of the form (\ref{eq:T_iso}) or (\ref{eq:T_non-iso}); in principle we could include single tensors with both inserted ancilla and post-selection, but we leave these out without loss of generality for the total map. All together, the tensors will constitute a linear map from the bulk legs (effective degrees of freedom $\ell r$) to the boundary legs on $\Lambda_\text{BH}$ (fundamental degrees of freedom $B$), and this map can itself be written as an isometry followed by post-selection:
\begin{equation} \label{eq:total_network}
    \prod_x T_x = \sqrt{|P'|} \langle0|_{P'} U' |0\rangle_{A'},
\end{equation}
where contractions between the different tensors $T_x$ in the product are performed according to the structure of the network and $U'$ acts on the whole system. Here, $A'$ represents all of the collective ancilla from the isometric tensors,
\begin{equation} \label{eq:A'}
    |0\rangle_{A'} = \bigotimes_x |0\rangle_A, \qquad \log_D |A'| = \sum_x \left( \#_{\text{out},x} - \#_{\text{in},x} \right) \Theta \left( \#_{\text{out},x} - \#_{\text{in},x} \right),
\end{equation}
where $\Theta$ is the Heaviside step function. Similarly, $P'$ represents all of the collective post-selection from the non-isometric tensors,
\begin{equation} \label{eq:P'}
    \langle0|_{P'} = \bigotimes_x \langle0|_P, \qquad \log_D |P'| = \sum_x \left( \#_{\text{in},x} - \#_{\text{out},x} \right) \Theta \left( \#_{\text{in},x} - \#_{\text{out},x} \right),
\end{equation}
delayed until the end of the network. $U'$ is then the collection of all $U_x$,
\begin{equation}
    U' = \prod_x U_x,
\end{equation}
where again contractions between different $U_x$ in the product are performed according to the structure of the network. 

Notice that the linear map given by the interior tensor network in (\ref{eq:total_network}) looks very similar to the generic holographic map $V\,:\, \ell r\rightarrow B$ defined by PHEVA in \cite{akers_black_2022}, 
\begin{equation} \label{eq:V_PHEVA}
    V = \sqrt{|P|} \langle\phi|_P U_H |\psi\rangle_f,
\end{equation}
where $U_H$ is a typical unitary drawn from the Haar measure acting on the entire Hilbert space $\hs_\ell \otimes \hs_r \otimes \hs_f$. Identifying $f$ with the inserted ancilla $A'$ in (\ref{eq:A'}) and $P$ with the post-selection $P'$ in (\ref{eq:P'}),\footnote{Any unitaries needed to relate $|0\rangle$ with $|\psi\rangle$ and $|\phi\rangle$ can be absorbed into the Haar random unitary $U_H$.} we see that the identification of (\ref{eq:total_network}) with (\ref{eq:V_PHEVA}) is complete if $U' = U_H$. However, requiring that $U'$ act on the full Hilbert space of $\ell rf$ as a Haar random unitary is a very strong condition. Instead, we will impose the more realistic requirement that $U'$ act as either a pseudorandom unitary \cite{ji_pseudorandom_2018} or unitary $k$-design on the Hilbert space of $\ell rf$.\footnote{There is a great deal of literature on constructing unitary $k$-designs from local random unitaries; see for example \cite{harrow_random_2009,brandao_local_2016,hunter-jones_unitary_2019}.} 

The Haar random properties of PHEVA's generic non-isometric map (\ref{eq:V_PHEVA}) were critical in demonstrating its usefulness on generic states of the effective description in \cite{akers_black_2022}. Because of it, the map was shown to act isometrically on average (with exponentially suppressed deviations), preserve subexponentially complex states, and give state-dependent reconstructions of interior operators. These features should then be viewed as requirements that any interior holographic map should satisfy. By restricting the unitaries $U_x$ used to construct the tensor network to act together as a pseudorandom or $k$-design unitary on the entire Hilbert space, the interior tensor network should inherit these same features. We will see in Sec.~\ref{sec:BFP} that the backwards-forwards-post-selection map -- proposed in \cite{dewolfe_non-isometric_2023} and shown in \cite{dewolfe_bulk_2024} to satisfy all requirements put forth in \cite{akers_black_2022} -- can be viewed as a particular example of these more generic interior tensor networks. This gives further evidence that an interior tensor network (\ref{eq:total_network}) can be constructed to satisfy the same properties.

Finally, we have provided a generalization of PHEVA's holographic map to include a notion of locality using hyperbolic tensor networks. We now turn to the properties of this new code, where we will find an interesting connection between non-isometries and quantum extremal surfaces (QESs).

\subsection{Structure of non-isometries in the inner code} \label{sec:noniso_structure}
\begin{figure}
    \centering
    \includegraphics[width=0.7\linewidth]{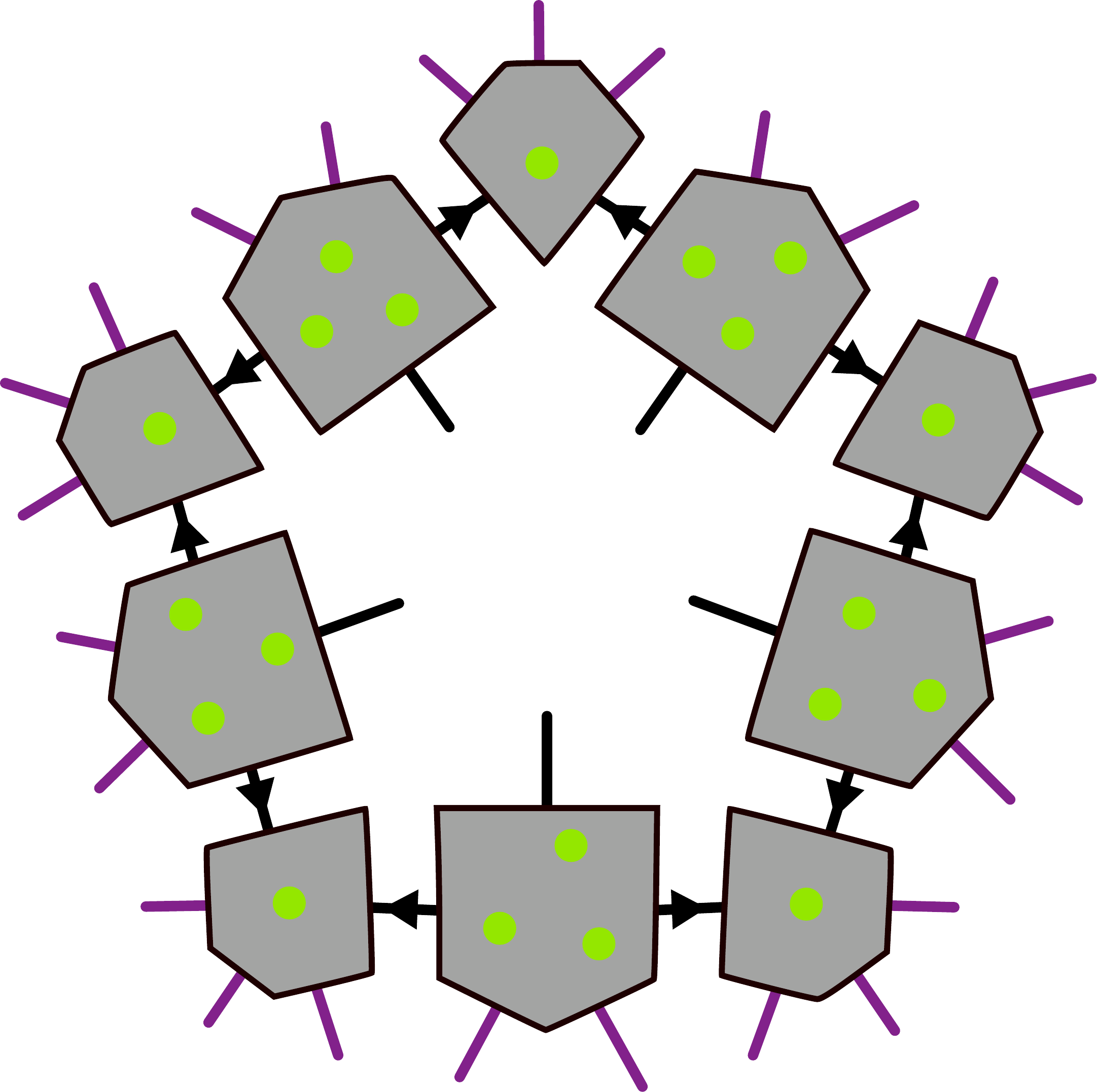}
    \caption{A depiction of the tensor network constructed from the first row ($N=1$) of a pentagonal tiling ($n=5$). Gray pentagons are the tensors $T_x$ associated with each cell of the tessellation, purple legs are uncontracted fundamental $B$ degrees of freedom, black legs are contracted between tensors, and green dots are bulk $r$ qudits. The inner contracted legs and all $r$ qudits are inputs to the row network, and the outer uncontracted legs are outputs. Legs contracted within the row are labeled with an arrow to indicate which are inputs or outputs to each tensor.}
    \label{fig:n5N1_unitary}
\end{figure}

Because our construction of the inner code incorporates a notion of locality and geometry, we have gained the ability to study the local structure of non-isometries within the code. We will continue implementing our assumption of spherically symmetric states, such that the placement of bulk qudits respects the same $2\pi/n$ symmetry as the tessellation. Our row parametrization will then be particularly useful here, and each row will be considered as one layer of the entire inner code. Taking all tensors in row $N$ together, the row provides a linear map from the uncontracted bulk legs in row $N$ and the contracted legs between rows $N-1$ and $N$ to the contracted legs between rows $N$ and $N+1$. Recalling that $q_B(N)$ defined in (\ref{eq:logB}) gives the number of legs on the outer edge of row $N$, the input size to row $N$ is
\begin{equation} \label{eq:input_size}
    \log_D\big|\,\text{input to row }N\,\big| = q_B(N-1) + q_r(N) + q_\ell(N),
\end{equation} 
while the output size is $q_B(N)$. Therefore, row $N$ will act non-isometrically if
\begin{equation} \label{eq:noniso_row_cond}
    q_r(N) + q_\ell(N) > q_B(N) - q_B(N-1), \qquad \text{non-isometric row}.
\end{equation}
Fig.~\ref{fig:n5N1_unitary} gives an example of the $N=1$ row of a $\{5,4\}$ hyperbolic tensor network where condition (\ref{eq:noniso_row_cond}) is exactly balanced and the row acts unitarily. If a single additional bulk $r$ qudit (represented as green dots) is added anywhere in the row, the entire row would act non-isometrically. We note that row $N$ could act non-isometrically if the bulk uncontracted legs are placed poorly, even if (\ref{eq:noniso_row_cond}) is not satisfied. Therefore satisfying (\ref{eq:noniso_row_cond}) guarantees non-isometry, but not satisfying it does not imply an isometric row.

The properties of a non-isometric row can vary depending on the amount of post-selection required. We will define a ``weakly non-isometric'' row as one in which the amount of post-selection is extremely small compared to the input size given by (\ref{eq:input_size}). The maps $V$ defined by such rows may behave as \textit{approximate isometries}, satisfying
\begin{equation} \label{eq:approx_isom}
    \| V^\dagger V - \mathbb{1} \|_\infty < \epsilon,
\end{equation}
for some small $\epsilon$. Inner products between the inputs to such a row would then be preserved to some high accuracy. However, if the amount of post-selection is some significant fraction of the input size, (\ref{eq:approx_isom}) is not satisfied and inner products are not even approximately preserved in general, indicating that information contained in the input to the row will be lost. Such maps will be referred to as ``strongly non-isometric''.

Just as we did for the rows, we can use qudit counting to get a sense of when the entire inner code is non-isometric. Suppose the inner code describes a black hole containing $N_0$ rows. The size of the input to the code is given by the total number of bulk qudits,
\begin{equation}
    \log_D\big|\,\text{input to code}\,\big| = \sum_{j=0}^{N_0} \Big( q_r(j) + q_\ell(j) \Big).
\end{equation}
If this exceeds the total number of fundamental $B$ qudits on the edge of the code,
\begin{equation} \label{eq:noniso_code_cond}
    \sum_{j=0}^{N_0} \Big( q_r(j) + q_\ell(j) \Big) > q_B(N_0), \qquad \text{non-isometric inner code}.
\end{equation}
then the inner code is guaranteed to be non-isometric. However, this counting neglects the fact that the inner code is defined as a composition of linear maps for each row. If a single row is non-isometric, the inner code will also be non-isometric automatically (since not all inner products will be preserved by the row in question) even if (\ref{eq:noniso_code_cond}) is not satisfied. Thus satisfying (\ref{eq:noniso_code_cond}) implies a non-isometric inner code, but not satisfying it does not imply isometry.

The strength of the inner code's non-isometry will be most directly correlated with the strength of non-isometries in the outermost rows of the tensor network, since the post-selection in these rows acts on nearly all bulk degrees of freedom. Thus only strongly non-isometric rows near the edge of the interior will likely lead to a strongly non-isometric code. Any other non-isometric rows -- such as those with only weak non-isometries or strongly non-isometric rows deep in the interior -- typically do not lead to a strongly non-isometric inner code. Therefore, if a strongly non-isometric row is present near the boundary of the interior, an approximately isometric encoding satisfying (\ref{eq:approx_isom}) is no longer possible. We note that while condition (\ref{eq:noniso_code_cond}) does not imply the presence of an outermost strongly non-isometric row, the condition is more likely to be satisfied if one is present.

While the non-isometric properties of individual rows and the entire inner code can be quite intricate, we have found that qudit counting can be useful in determining general properties. The change in the number of $B$ qudits between two rows was particularly important in this analysis, so it will be useful to understand the typical change in $q_B(N)$. We note that this change is simplest for flat tessellations, where $q_B(N)$ reduces to $n(1+2N)$; thus we find
\begin{equation}
    q_B(N) - q_B(N-1) = 2n, \qquad \text{flat tessellation}.
\end{equation}
The change in the number of $B$ qudits is constant for flat tessellations. Unfortunately, there is no similarly compact expression for the change in $q_B(N)$ for hyperbolic tessellations. Instead, we can find an approximate expression by taking the $N\rightarrow\infty$ limit of a fractional change across row $N$:
\begin{equation} \label{eq:largeN}
    \lim_{N\rightarrow\infty} \frac{q_B(N) - q_B(N-1)}{q_B(N)} = 2 \left( 1 + \sqrt{\frac{(n-2)(k-2)}{k(n-2)-2n}} \, \right)^{-1}
\end{equation}
This limit yields a fractional loss of $0$ for flat tessellations; this does not disagree with the exact result above since there is a $1/N$ dependence lost in the large $N$ limit. For hyperbolic tessellations, the fractional change is smallest for a $\{3,7\}$ tessellation, with a value of $\sim0.62$. Therefore, any interior tensor network built from a hyperbolic lattice will lose more than $50\%$ of its $B$ qudits when it is shrunk by one row.

\subsection{Quantum extremal surfaces} \label{sec:QES}
Similarly, the incorporation of geometry into this qudit black hole model through tensor networks allows us to identify quantum extremal surfaces (QESs) in the model. Candidate QESs will be identified as local minima of the generalized entropy,
\begin{equation}
    S_\text{gen}(\chi) = \frac{A_\chi}{4G_N} + S(EW),
\end{equation}
where $\chi$ is a codimension-2 surface homologous to a boundary subregion and $S(EW)$ is the entropy of any bulk fields in the entanglement wedge defined by $\chi$ and the boundary. A similar generalized entropy can be defined for tensor networks; following the work of \cite{pastawski_holographic_2015,hayden_holographic_2016} and taking all logarithms to be base $D$, we replace the $A/4G_N$ term with the number of legs $|\chi|$ cut by $\chi$,
\begin{equation}
    S_\text{gen}(\chi) = |\chi| + S(EW).
\end{equation}
$S(EW)$ will be given by the entropy of all bulk qudits in the entanglement wedge bounded by $\chi$.

We are particularly interested in identifying QESs for the generalized entropy of the black hole $B$ and its radiation $R_\text{out}$, so we will consider surfaces $\chi$ that are homologous to the entire boundary. (Since all degrees of freedom outside $\Lambda_\text{BH}$ have been removed to the reservoir $R$, we take $\Lambda_\text{BH}$ to be the boundary of the interior tensor network and use it for the homology constraint.) For simplicity, we will use our assumption of spherically symmetric states to restrict to surfaces $\chi$ that are similarly symmetric, with each living along the outside edge of a row. We will then define a surface $\chi$ by the number of tessellation rows it contains. The area $|\chi|$ of the surface containing $N$ rows is then given by the number of contracted legs between row $N$ and row $N+1$, equivalent to $q_B(N)$ defined in (\ref{eq:logB}). 

To calculate the entropy of bulk degrees of freedom $S(EW)$, we recall that all interior bulk degrees of freedom are categorized as either $r$ or $\ell$. Radiation $r$ qudits are maximally entangled with radiation $R_\text{out}$ in the reservoir, so each unpaired radiation qudit contributes $\log_D D = 1$ to $S(EW)$. For simplicity, let us assume that each $\ell$ qudit caries the same entanglement 
\begin{equation}
    s = \frac{S(R_\text{in})}{\log_D|\ell|}
\end{equation}
with some reference. Therefore, each $\ell$ qudit contributes $0\leq s \leq1$ to $S(EW)$.

Now, consider a black hole containing $N_0$ rows. The generalized entropy of the black hole $S_{\text{gen},B}$ makes use of the entanglement wedge stretching between $\chi$ and the boundary. For a surface $\chi$ on the outside edge of row $N$, the generalized entropy is
\begin{equation}
    S_{\text{gen},B} (N) = q_B(N) + \sum_{j=N+1}^{N_0} \Big(q_r(j) + sq_\ell(j)\Big),
\end{equation}
and the generalized entropy of $B$ is given by the minimization over $N$. Similarly, the generalized entropy of the radiation $R_\text{out}$ makes use of the complimentary entanglement wedge containing the reservoir and the island behind $\chi$,
\begin{equation}
    S_{\text{gen},R_\text{out}} (N) = q_B(N) + \sum_{j=N+1}^{N_0} q_r(j) + \sum_{k=0}^N sq_\ell(k),
\end{equation}
where again the generalized entropy of $R_\text{out}$ is given by the minimization over $N$.

Candidate QESs are given by local minima of these generalized entropies. We therefore define a ``local'' change $\delta S_\text{gen}$ in generalized entropy from row $N-1$ to row $N$ as
\begin{equation}
    \delta S_\text{gen} (N) = S_\text{gen}(N) - S_\text{gen}(N-1).
\end{equation}
For the generalized entropy of $B$ and $R_\text{out}$, this local change is given by
\begin{align}
    \delta S_{\text{gen},B} (N) &= q_B(N) - q_B(N-1) - q_r(N) - sq_\ell(N) \\
    \delta S_{\text{gen},R_\text{out}} (N) &= q_B(N) - q_B(N-1) - q_r(N) + sq_\ell(N) \label{eq:local_Srout}
\end{align}
We note that both only depend on the bulk degrees of freedom in row $N$ and only differ in the sign of the contribution from the $\ell$ degrees of freedom. Using either formula, a candidate QES will be identified as $\delta S_\text{gen} (N) \leq 0$; a candidate QES at non-zero $N$ will be referred to as a ``non-trivial'' QES. For $s=0$, this condition for the existence of a candidate QES reduces to the same condition for both $B$ and $R_\text{out}$:
\begin{equation} \label{eq:QES_cond}
    q_r(N) \geq q_B(N) - q_B(N-1).
\end{equation}
Therefore, if the number of $r$ legs in row $N$ exceeds the difference between the number of contracted legs leaving row $N$ and the number entering row $N$ from row $N-1$, a candidate QES will form in row $N$. Interestingly, this is the same as condition (\ref{eq:noniso_row_cond}) for row $N$ to be non-isometric with $q_\ell(N)$ or $s$ set to zero! A QES at row $N$ satisfying (\ref{eq:QES_cond}) will then imply that (\ref{eq:noniso_row_cond}) is satisfied, no matter the value of $q_\ell(N)$. Therefore, the presence of a non-trivial QES implies the row must be non-isometric, although the lack of one does not imply isometry.

If there are multiple candidate QESs, we can compare their generalized entropy to determine which is the true global minimum. In the case of an evaporating black hole, we expect two candidate QESs: the zero-area surface at row $N=0$ and a non-trivial QES at row $N\neq0$. We therefore define a ``global'' change $\Delta S_\text{gen}$ in generalized entropy from row 0 to row $N$ as
\begin{equation}
    \Delta S_\text{gen} (N) = S_\text{gen} (N) - S_\text{gen} (0).
\end{equation}
For the generalized entropy of $B$ and $R_\text{out}$, this global change is given by
\begin{align}
    \Delta S_{\text{gen},B} (N) &= q_B(N) + \sum_{j=0}^{N} \Big(-q_r(j) - sq_\ell(j)\Big) \\
    \Delta S_{\text{gen},R_\text{out}} (N) &= q_B(N) + \sum_{j=0}^{N} \Big(-q_r(j) + sq_\ell(j)\Big) \label{eq:global_Rout}
\end{align}
The non-trivial QES will become the global minimum when $\Delta S_\text{gen} (N) \leq 0$. For $s=0$, when the infalling matter creating the black hole is in a pure state, this reduces to the same condition for both global changes:
\begin{equation}
    \sum_{j=0}^N q_r(j) \geq q_B(N).
\end{equation}
Therefore, if the total number of $r$ legs behind row $N$ exceeds the number of contracted legs leaving row $N$, the non-trivial QES will be the global minimum. Interestingly, this is the same as
the naive condition (\ref{eq:noniso_code_cond}) for the inner code comprised of rows $0$ to $N$ to act non-isometrically on all bulk degrees of freedom in those rows, again with $q_\ell(N)$ or $s$ set to zero. While we already know that the map formed by rows $0$ through $N$ acts non-isometrically (due to the non-isometry of row $N$ indicated by $\delta S_\text{gen} \leq 0$), a globally minimal QES satisfying $\Delta S_\text{gen} \leq 0$ is more likely to indicate that row $N$ is strongly non-isometric as described in the previous subsection. In this case, the entire map formed by rows 0 through $N$ is likely strongly non-isometric as well and cannot be described by an approximate isometry.

The addition of locality to these non-isometric black hole codes has bought us a powerful tool. Not only are we able to identify and track QESs in the model, but we have shown a strong correspondence between the existence of non-trivial QESs and the structure of non-isometries in the holographic code.

\subsection{Interior operator reconstruction} \label{sec:op_recon}
Finally, let us consider how interior operators encoded by this interior tensor network could be reconstructed by an external observer. We make no new claims about the ability to reconstruct operators behind the horizon, but it is important to see how expectations from previous results are realized here.

First, consider a tensor network holographic map for an AdS bulk without a black hole. There is no horizon, no inner code, and no non-trivial QESs. The outer code acts as an exact isometry, and bulk operators are encoded redundantly on the boundary. All reconstructions of the operator act in the same way on the code subspace, with each reconstruction corresponding to different entanglement wedges that contain the bulk operator.

However, we expect from \cite{penington_entanglement_2020,almheiri_entropy_2019} that the existence of a black hole will introduce one non-trivial QES in the outermost row of the black hole interior (just behind the horizon). If we consider an evaporating black hole, the Page time will play an important role. Long before the Page time, the non-trivial QES should not be the global minimum. 
Thus there will be enough bulk degrees of freedom to form a QES in the outermost row, but not enough to outnumber the fundamental $B$ degrees of freedom. In a large black hole with a large number of tensors, such a QES is likely to correspond to a weakly non-isometric outermost row. 

It is then possible for the resulting inner code to behave as an approximate isometry, preserving inner products of all states up to some small error without the need for averaging. In appendix D of \cite{akers_black_2022}, PHEVA found that their generic holographic map (\ref{eq:V_PHEVA}) was approximately isometric when
\begin{equation} \label{eq:approx_isom_cond}
    \frac{|\ell||r|}{|B|} |B|^{2\gamma} \left( \gamma\log_D|B| + \log_D|P| \right) \ll 1, \quad 0 < \gamma < 1/2.
\end{equation}
We expect a similar requirement should be true of our tensor network inner code so long as the above requirement that $U'$ in (\ref{eq:total_network}) acts as a pseudorandom unitary or unitary $k$-design holds. In our case, $\log_D|B|$ would be given by (\ref{eq:logB}), $\log_D|P|$ by (\ref{eq:P'}), and $\log_D|\ell||r|$ by the number of bulk legs in the interior tensor network. 

So long as (\ref{eq:approx_isom_cond}) is true, the tensor network inner code will act as an approximate isometry on all bulk states. Operators acting on both $\ell$ and $r$ in the interior will be encoded in $B$ up to some small error, and the outer code will encode these operators in the CFT degrees of freedom on the boundary. Because the entire holographic code acts as an approximate isometry, a boundary observer will be able to reconstruct an interior operator (to some precision) from all boundary data. However, the process of reconstructing the operator will be exponentially complex due to the presence of post-selection in the inner code at the location of the non-trivial QES. This then gives an explicit construction of the python's lunch proposal \cite{brown_pythons_2020}, with the non-isometric tensors in the row containing the QES forming the mouth of the lunch.

We pause to note that because $\delta S_\text{gen} < 0$ at the non-trivial QES, the greedy algorithm introduced in \cite{pastawski_holographic_2015} to find minimal surfaces would fail to capture the outermost row of the inner code due to the lack of fundamental legs $B$ on the boundary of the inner code. While the greedy algorithm locates the non-trivial QES, it does not know that it is not the global minimum and would fail to demonstrate the reconstructability of operators behind the horizon before the Page time.

At the Page time, we expect the non-trivial QES will transition to become the global minimum. We understand from the previous subsections that this indicates the inner code will map a large number of bulk degrees of freedom $\ell r$ to a comparatively small number of fundamental degrees of freedom $B$. Furthermore, the black hole would have shrunk to a smaller size, forcing the large number of bulk degrees of freedom into a smaller number of tensors. This suggests that the globally minimal QES represents a strongly non-isometric outermost row. The entire inner code will then be strongly non-isometric as well and unable to act as an approximate isometry.

In this regime, the interior tensor network will no longer be capable of encoding interior operators in $B$ or the corresponding CFT degrees of freedom. No amount of boundary data will allow a boundary observer to reconstruct interior operators to some high precision. Instead, we should find that interior operators are reconstructable on $R_\text{out}$, allowing an observer with access to all of the radiation to reconstruct interior operators. 

Thus two tasks remain. First, we need to verify that these QES dynamics are reproduced in this tensor network model of the interior. Second, we must demonstrate reconstructability of interior operators in the radiation $R_\text{out}$ long after the Page time. To accomplish both, we must specify dynamics for bulk and boundary degrees of freedom in this qudit model of an evaporating black hole. In the next section we describe a possible model of dynamics based on those used in \cite{akers_black_2022,dewolfe_non-isometric_2023,dewolfe_bulk_2024}, viewing the model as a prescription for assigning degrees of freedom to each cell of a hyperbolic tessellation.

\section{Time evolution and the inner code} \label{sec:dynamics}

In the previous section, we used hyperbolic tensor networks to add a notion of locality to the non-isometric holographic maps proposed by PHEVA in \cite{akers_black_2022}. Most notably, the added geometry gives us the ability to identify quantum extremal surfaces (QESs) in these qudit models of the black hole interior. QESs are known to behave in very specific ways for evaporating black holes \cite{penington_entanglement_2020,almheiri_entropy_2019}, and these QES dynamics have implications for the reconstructability of interior operators; see Sec.~\ref{sec:op_recon}. In this section, we aim to introduce dynamics for an evaporating black hole so that we can check that our model realizes the correct QES behavior.

Before describing time evolution of the bulk effective and boundary fundamental degrees of freedom, we pause to make a few comments about time evolution in tensor networks. Constructing a full tensor network model of AdS/CFT with dynamics is an open problem \cite{faist_continuous_2020,dolev_gauging_2022}. Dynamics in both the AdS bulk and CFT boundary should be local; this is a requirement imposed by both theories. However, it is well known that tensor network representations of holographic dualities encode bulk operators non-locally in the boundary. Therefore, local dynamics deep in the bulk would be dual to very non-local dynamics in the boundary, appearing to violate the locality of CFT dynamics.

There have been several interesting attempts to remedy this issue. Work by Apel, Kohler, and Cubitt has used Hamiltonian simulation to create a local boundary Hamiltonian that models the non-local boundary Hamiltonian dual to a local bulk Hamiltonian \cite{kohler_toy_2019,apel_holographic_2022,apel_security_2024}. Osborne and Stiegemann have studied Thompson's group as a discrete version of conformal transformations of the boundary, finding that unitary representations of Thompson's group are dual to transformations of triangular hyperbolic tessellations \cite{osborne_dynamics_2020}. Finally, tensor networks with additional gauge symmetries are being pursued as a means of creating a more physical bulk theory that might permit a local boundary Hamiltonian \cite{donnelly_living_2017,qi_emergent_2022,dolev_gauging_2022,dong_holographic_2024,akers_background_2024,akers_multipartite_2024}.

Since there is still much to learn about the dynamics of tensor networks, we will limit our dynamics to the movement of bulk degrees of freedom between the cells of a fixed tessellation. We will not attempt to describe dynamics of the tessellation (and resulting tensor network structure) itself. We then view these dynamics as a prescription for assigning bulk degrees of freedom to the cells of a tessellation in a way consistent with the evolution of an evaporating black hole.

As we move bulk degrees of freedom around a tessellation, we will use unitaries to describe interactions between them. We will be as general as possible in the descriptions of these unitaries, similar to the unitaries described in \cite{dewolfe_non-isometric_2023,dewolfe_bulk_2024}. Furthermore, we will not attempt to make use of any duality between the dynamics of the effective and fundamental descriptions, even though one should in principle be present. Our hope is that further details can be filled in later once a better understanding of tensor network dynamics is reached. It might even be possible that future work comparing the dynamical and non-dynamical inner codes we describe in this work could constrain the duality between the bulk and boundary dynamics.

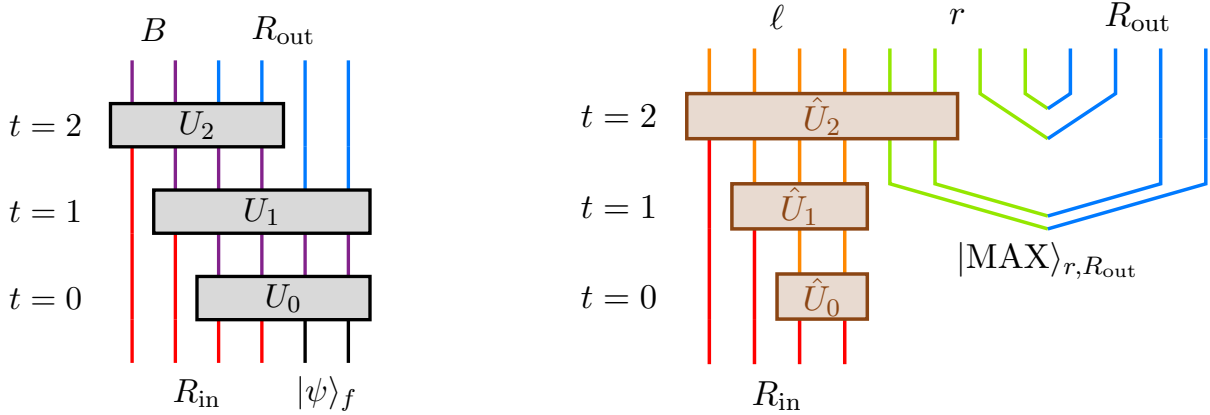
\begin{figure}
    \centering
    \subfloat{\input{fun_dyn}}
    \hfill
    \subfloat{\input{eff_dyn}}
    \caption{Circuit diagrams representing the fundamental dynamics $U_t$ (left) and \textcolor{brown}{effective dynamics} $\hat{U}_t$ (right) until $t=2$ for $n_0=4$, $m_0=2$, $\cI=1$, and $\cO=2$. These are used as the basis for the dynamics of degrees of freedom localized in hyperbolic tessellations. Lines are colored by qudit type according to Fig.~\ref{fig:colors}: \textcolor{red}{red} for $R_\text{in}$, \textcolor{blue}{blue} for $R_\text{out}$, \textcolor{purple}{purple} for $B$, \textcolor{orange}{orange} for $\ell$, \textcolor{green}{green} for $r$, and black for $f$. Figure reproduced from \cite{dewolfe_bulk_2024}.}
    \label{fig:fun_eff_dyn}
\end{figure}

With these preliminaries out of the way, we give descriptions for the dynamics in both the fundamental and effective descriptions. These are based on the dynamics described in \cite{akers_black_2022,dewolfe_non-isometric_2023,dewolfe_bulk_2024}; the corresponding circuit diagrams are reproduced from \cite{dewolfe_bulk_2024} here in Fig.~\ref{fig:fun_eff_dyn}. The goal of the following prescriptions will be to add a notion of locality through hyperbolic tessellations to these dynamics.

\subsection*{Fundamental dynamics}
We begin with the dynamics of the fundamental description. Initially at time $t=0$, $m_0$ qudits from $R_\text{in}$ and $n_0 - m_0$ qudits in a fixed state $|\psi\rangle_f$ are acted on by a unitary $U_0$ and placed along the outer boundary of the $N_0^\text{th}$ row of the tessellation. Equation (\ref{eq:logB}) must relate $n_0$ to the number of edges on row $N_0$ as $n_0 = q_B(N_0)$ so that only one qudit is placed on each edge of the tessellation along $\Lambda_\text{BH}$. This defines the initial state of our fundamental $B$ degrees of freedom.

At each following time step, $\cI$ qudits from $R_\text{in}$ in the reservoir are taken and added to $B$, a unitary $U_t$ is applied to all $B$ qudits (including the additional $\cI$ qudits), and then $\cO$ are released as radiation $R_\text{out}$ and removed to the reservoir.\footnote{Realistically, $\cO$ should increase with decreasing $N$, since the temperature of a black hole increases with decreasing radius. We will leave $\cO$ general in this work.} All unitaries $U_t$ are taken to be random (either pseudorandom or unitary $k$-designs) and all-to-all.  The change in the size of $B$ at each time step $\Delta t$ is then given by
\begin{equation}
    \frac{\Delta\log_D|B|}{\Delta t} = \cI - \cO.
\end{equation}
Evaporation of the black hole then requires $\cI < \cO$. This describes the ``fine-grained'' time evolution of the fundamental description and is depicted as a circuit diagram (without the tessellation) in the left panel of Fig.~\ref{fig:fun_eff_dyn}.

\begin{figure}
    \centering
    \includegraphics[width=\linewidth]{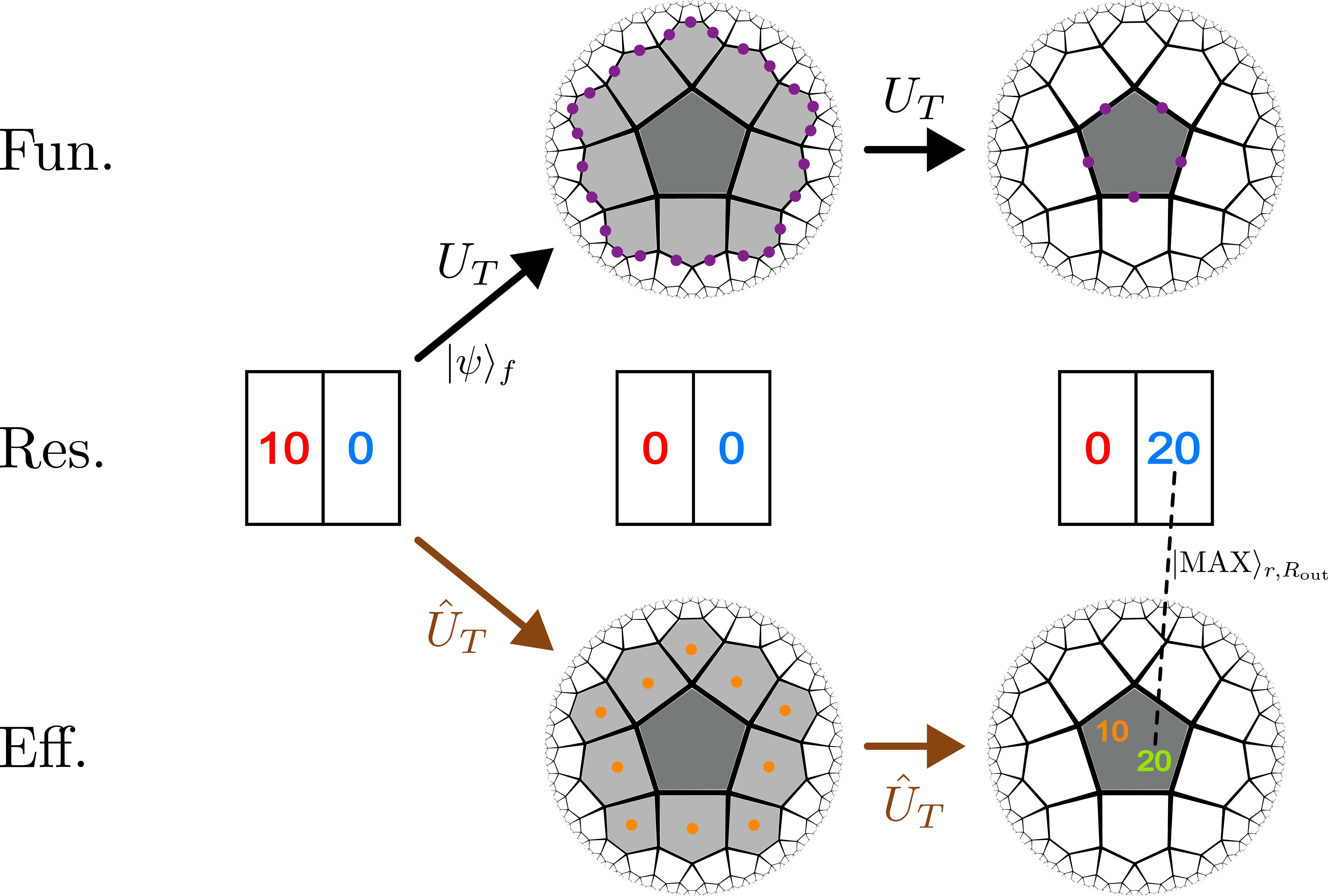}
    \caption{The dynamics of a black hole initially containing two rows in a $\{5,4\}$ hyperbolic tessellation. Colored cells indicate the part of the tessellation behind $\Lambda_\text{BH}$. One dot represents one qudit, numbers represent multiple qudits; both dots and numbers are colored by qudit type. The upper sequence describes evolution in the fundamental description, while the lower sequence describes evolution in the effective description. The reservoir $R$ is depicted between the two sequences and is shared by both descriptions. $U_T$ and $\hat{U}_T$ are used to represent $U_{t+\Delta T}\dots U_{t+1} U_t$ and $\hat{U}_{t+\Delta T}\dots \hat{U}_{t+1} \hat{U}_t$, respectively.}
    \label{fig:dynamics_N2}
\end{figure}

Because our tensor network model for the interior is discrete, it can only describe black holes of discrete size $\log_D|B| = q_B(N)$ given by (\ref{eq:logB}). As a black hole shrinks in this model, it does so by ``shedding'' its outermost row behind $\Lambda_\text{BH}$; this corresponds to a change in the size of $B$ given by
\begin{equation}
    \Delta \log_D|B| = q_B(N) - q_B(N-1).
\end{equation}
Anything less would require changes to the tessellation and network,\footnote{In some cases, shedding individual cells could give a smaller change in $\log_D|B|$. However, this would require us to abandon our assumption of angular symmetry, so we will not consider such dynamics here.} which we are not considering in our model of time evolution. We denote the total number of time steps $\Delta t$ required for an evaporating black hole to shed row $N$ as $\Delta T(N)$, given by
\begin{equation} \label{eq:dT}
    \Delta T (N) = \frac{q_B(N) - q_B(N-1)}{\cO - \cI}.
\end{equation}
After this time, the remaining $B$ qudits (now of size $q_B(N-1)$) are placed along the outer edges of row $N-1$. Since $\Delta T > \Delta t$, the dynamics of an evaporating black hole in this tensor network model (given by shedding rows) is a ``coarse-grained'' version of the dynamics described above. The top sequence of Fig.~\ref{fig:dynamics_N2} gives an example of the (coarse-grained) dynamics of the fundamental description for a $\{5,4\}$ tessellation with $m_0 = 10$, $N_0 = 2$, and $\cI=0$. The reservoir $R$ is depicted in the middle of the diagram as a box containing the number of qudits in $R_\text{in}$ and $R_\text{out}$.

\subsection*{Effective dynamics}
Let us now turn to the dynamics of the effective description. Initially at $t=0$, the same $m_0$ qudits from $R_\text{in}$ are acted on by a unitary $\hat{U}_0$; the resulting $m_0$ qudits are taken from the reservoir, relabeled as $\ell$ degrees of freedom, and placed in the cells of the outermost row of a hyperbolic tessellation with $N_0$ rows.\footnote{In principle, $m_0$ and $N_0$ should be related by the energy density required to form a black hole of radius $O(N_0 L)$. In the example dynamics shown in Fig.~\ref{fig:dynamics_N2}, we have set $m_0$ equal to the number of cells in row $N_0$, but this was just chosen to give a nice example.} This defines the initial state of the black hole interior in the effective description.

At each future time step $\Delta t$,
\begin{enumerate}
    \item $\cI$ qudits from $R_\text{in}$ are taken from the reservoir and inserted into the outermost row of cells behind $\Lambda_\text{BH}$ as new $\ell$ qudits.
    \item A unitary $\hat{U}_t$ is applied to all qudits in the interior of the black hole.
    \item $\cO$ qudits of $r$ and $R_\text{out}$ are inserted in the maximally entangled state,
        \begin{equation}
            |\text{MAX}\rangle_{r,R_\text{out}} = \frac{1}{\sqrt{D}} \sum_{j=1}^D |j\rangle_r |j\rangle_{R_\text{out}},
        \end{equation}
    $r$ is inserted into the outermost row of the interior, while $R_\text{out}$ is inserted into the reservoir.
\end{enumerate}
Each $\hat{U}_t$ should be constructed in a way that respects the locality of the hyperbolic tessellation. For example, $\hat{U}_t$ could be constructed from local operators that only act on neighboring cells, coupling all of the qudits in one cell with all of the qudits in an adjacent cell. Longer range interactions could also be included by coupling two cells with an interaction that decreases in strength with the separation of the cells. We make no particular choice here, but we assume that $\hat{U}_t$ is local in this sense. Furthermore, we assume that the initial placement of new qudits and the local structure of $\hat{U}_t$ respect the $2\pi/n$ rotational symmetry of the tessellation. All together, this gives the ``fine-grained'' dynamics of the effective description as depicted (without the locality given by the tessellation) in the right panel of Fig.~\ref{fig:fun_eff_dyn}.

Again, we must ``coarse-grain'' these dynamics to make them consistent with the discrete nature of the tessellation. After $\Delta T(N)$ time steps -- as given by equation (\ref{eq:dT}) -- the black hole sheds one row. The total number of bulk qudits added to the interior of the black hole during time $\Delta T(N)$ is
\begin{align}
    \Delta \log_D|r| &= \frac{\cO}{\cO - \cI} \big( q_B(N) - q_B(N-1) \big) \label{eq:bulkDOFs_perRow} \\
    \Delta \log_D|\ell| &= \frac{\cI}{\cO - \cI} \big( q_B(N) - q_B(N-1) \big) 
\end{align} 
All interior degrees of freedom (both $\ell$ and $r$) must remain within the interior, so any bulk qudits in the outermost row must move in at least one row when the outermost row is shed. The bottom sequence of Fig.~\ref{fig:dynamics_N2} gives an example of the (coarse-grained) effective description dynamics for a $\{5,4\}$ tessellation with $m_0 = 10$, $N_0 = 2$, and $\cI=0$. The reservoir in the middle row is shared by both descriptions.

We note that outside of requiring all that qudits $\ell r$ remain within the interior when the black hole sheds a row, we have been purposefully vague about the movements of bulk qudits within the interior. Determining the movement of a qudit between cells would require knowledge of its velocity and interactions with other qudits. Since we are leaving $\hat{U}_t$ general, we do not attempt to model the specific movements of bulk degrees of freedom here.

\subsection{Wormholes in dynamically generated states} \label{sec:wormhole}
In the above dynamics for the effective bulk description, all Hawking pairs $rR_\text{out}$ are initially in the maximally entangled state of two qudits. This necessarily restricts the set of states accessible by the effective dynamics to a subspace of size $|R_\text{in}|$ of the full effective Hilbert space; we refer to this as the ``dynamically generated subspace''. Let us now restrict ourselves to this subspace and consider how our interior tensor network construction behaves on such states. Remarkably, we will find wormholes forming from the insertion of the maximally entangled Hawking pairs!

We will find the PEPS notation for tensor networks \cite{verstraete_renormalization_2004} very useful in this analysis. We provide a short review here; see \cite{hayden_holographic_2016,akers_background_2024} for more thorough reviews of the PEPS notation. In this framework, a tensor $T$ with $m$ legs is used to define a state in a product Hilbert space as
\begin{equation} \label{eq:T_as_product}
    |T\rangle = \sum_{\{\mu\}} T_{\mu_1 \dots \mu_m} |\mu_1\rangle\otimes\dots\otimes|\mu_m\rangle,
\end{equation}
where $|\mu_i\rangle$ forms an orthonormal basis for a Hilbert space associated to the $i^\text{th}$ leg of the tensor. For every vertex $x$ in a network, we associate one tensor state $|T_x\rangle$, the collection forming a product state $\bigotimes_x |T_x\rangle$. Contracting legs of the tensors along the links of the network is done by post-selecting on the maximally entangled state between two vertices,
\begin{equation} \label{eq:contract}
    |\text{TN}\rangle = \Bigg( \bigotimes_{\langle xy \rangle} \, \langle\text{MAX}|_{x,y} \Bigg) \Bigg( \bigotimes_x |T_x\rangle \Bigg),
\end{equation}
to sum over the contractions. Here we have used angle brackets $\langle\dots\rangle$ to denote neighboring vertices. A state of the effective bulk description $|\hat{\Psi}\rangle\in\hs_\text{eff}$ can be fed into the tensor network by post-selecting on a subset of the remaining free legs using its conjugate $\langle\hat{\Psi}|$,\footnote{An equivalent PEPS representation of the holographic tensor network is found by taking the adjoint of (\ref{eq:PEPS_map}). In this case, the network is formed by tensor states $\langle T_x|$ living in the dual Hilbert space with contracted legs given by maximally entangled ket states $|\text{MAX}\rangle_{x,y}$. This has the advantage of making the holographic map look like an operator that acts on an input state $|\hat{\Psi}\rangle$ from the left. However, the output of the tensor network $\langle\Psi|$ lives in the dual Hilbert space. All results in this work using the PEPS notation can be converted to this second viewpoint by taking the adjoint of any expression.}
\begin{equation} \label{eq:PEPS_map}
    |\Psi\rangle = \langle\hat{\Psi}| \Bigg( \bigotimes_{\langle xy \rangle} \, \langle\text{MAX}|_{x,y} \Bigg) \Bigg( \bigotimes_x |T_x\rangle \Bigg),
\end{equation}
where the resulting output is a state of the fundamental description $|\Psi\rangle\in\hs_\text{fun}$.

Let us now specialize to hyperbolic tensor networks constructed from the dynamically generated subspace of the interior. At a particular moment in time, we have a hyperbolic tessellation $\{n,k\}$ of the black hole interior behind $\Lambda_\text{BH}$ with some number of $\ell$ and $r$ qudits associated to each cell according to the effective dynamics above. Similar to the construction in Sec.~\ref{sec:TN}, we first associate a tensor state $|T_x\rangle$ to each cell of the tessellation; this tensor must have enough legs (equivalently, enough factors $|\mu_m\rangle$ in $|T_x\rangle$) to accommodate both the bulk degrees of freedom associated to that cell and the $n$ legs needed to contract with neighboring cells. The tensor network is completed by post-selecting with $\langle\text{MAX}|_{x,y}$ to contract legs between neighboring cells $\langle xy\rangle$ of the hyperbolic tessellation. The result is a tensor network that maps bulk effective degrees of freedom $\ell r$ associated with each cell to fundamental degrees of freedom $B$ living on the legs dangling across $\Lambda_\text{BH}$ at the boundary of the network.

\begin{figure}
    \centering
    \includegraphics[width=0.6\linewidth]{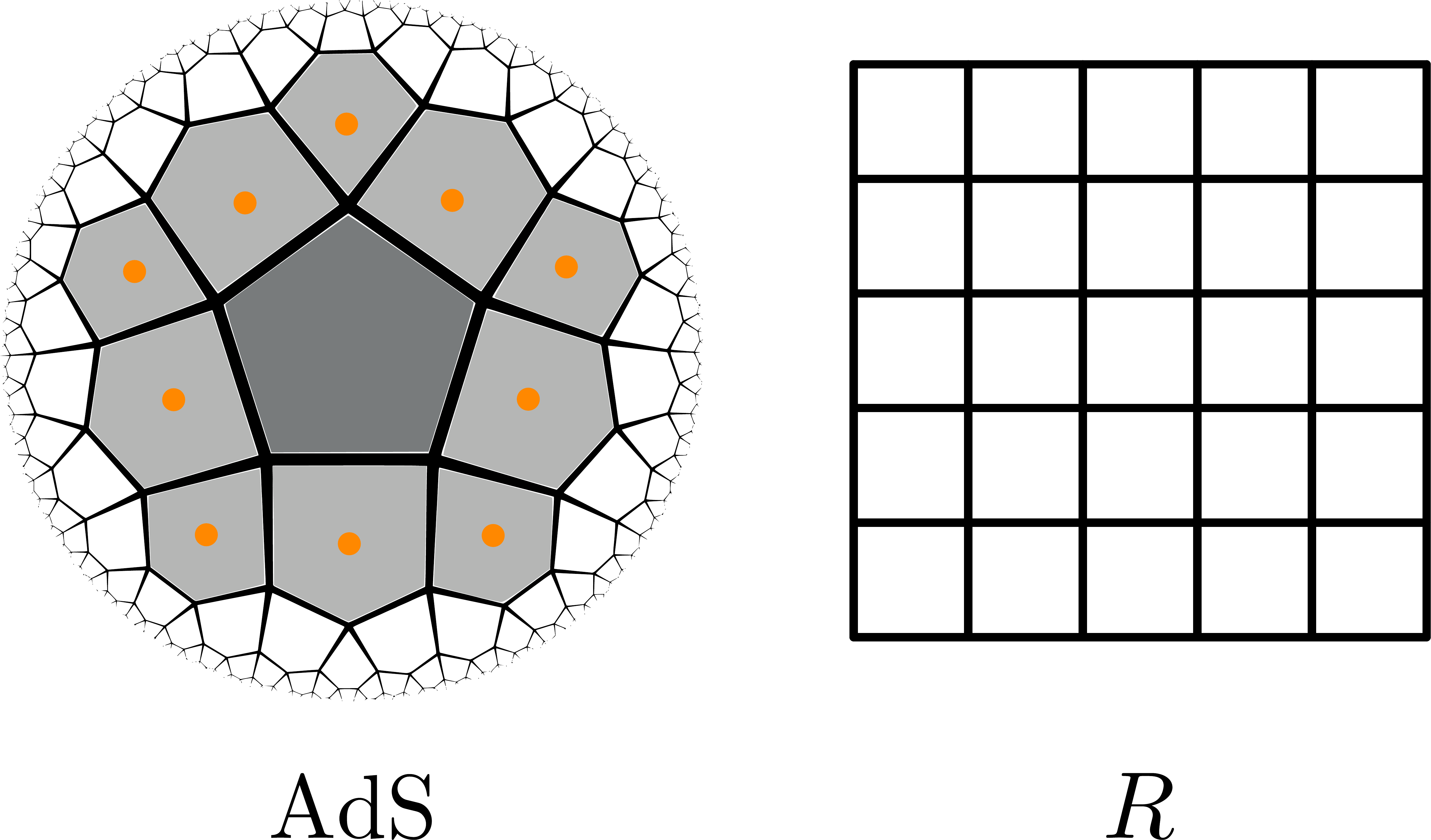}
    \caption{Tessellations and corresponding bulk degrees of freedom for an AdS black hole (left) and flat reservoir (right) for $N_0 = 2$ and $m_0 = 10$ at $t=0$. Colored cells indicate the part of the AdS black hole tessellation behind $\Lambda_\text{BH}$, and dots represent qudits colored according to Fig.~\ref{fig:colors}.}
    \label{fig:AdS_R_Tess}
\end{figure}

So far we have left the specifics of the reservoir ambiguous, as any non-dynamical spacetime would be suitable to hold the external degrees of freedom. Just as with the AdS black hole interior, this space can be tessellated and filled with a tensor network. For example, we might choose the reservoir to be flat space and tessellate it with a $\{4,4\}$ tessellation. Fig.~\ref{fig:AdS_R_Tess} shows an example of this configuration of black hole and reservoir for $N_0 = 2$ and $m_0 = 10$ at $t=0$. Given this tessellation, we can create a tensor network for the reservoir as well; the combined tensor networks for the AdS black hole interior and reservoir are then
\begin{equation} \label{eq:AdS_R_PEPS}
    |\text{TN}\rangle_\text{AdS} \otimes |\text{TN}\rangle_R = \Bigg( \bigotimes_{\langle xy \rangle} \, \langle\text{MAX}|_{x,y} \Bigg) \Bigg( \bigotimes_x |T_x\rangle_\text{AdS} \Bigg) \otimes \Bigg( \bigotimes_{\langle xy \rangle} \, \langle\text{MAX}|_{x,y} \Bigg) \Bigg( \bigotimes_x |T_x\rangle_R \Bigg).
\end{equation}
We note that the AdS and $R$ spacetimes are disconnected, so their respective tensor networks $|\text{TN}\rangle_\text{AdS}$ and $|\text{TN}\rangle_R$ are separable as a tensor product. Equation (\ref{eq:AdS_R_PEPS}) provides a map into which we can feed effective bulk degrees of freedom $\langle\hat{\Psi}|$ at any time step of the effective evolution and receive fundamental degrees of freedom $|\Psi\rangle$ at the same time step of the fundamental description. The only change in (\ref{eq:AdS_R_PEPS}) itself between time steps will be in the dimension of each $|T_x\rangle$ to accommodate the changing number of bulk inputs.

\begin{figure}
    \centering
    \includegraphics[width=0.6\linewidth]{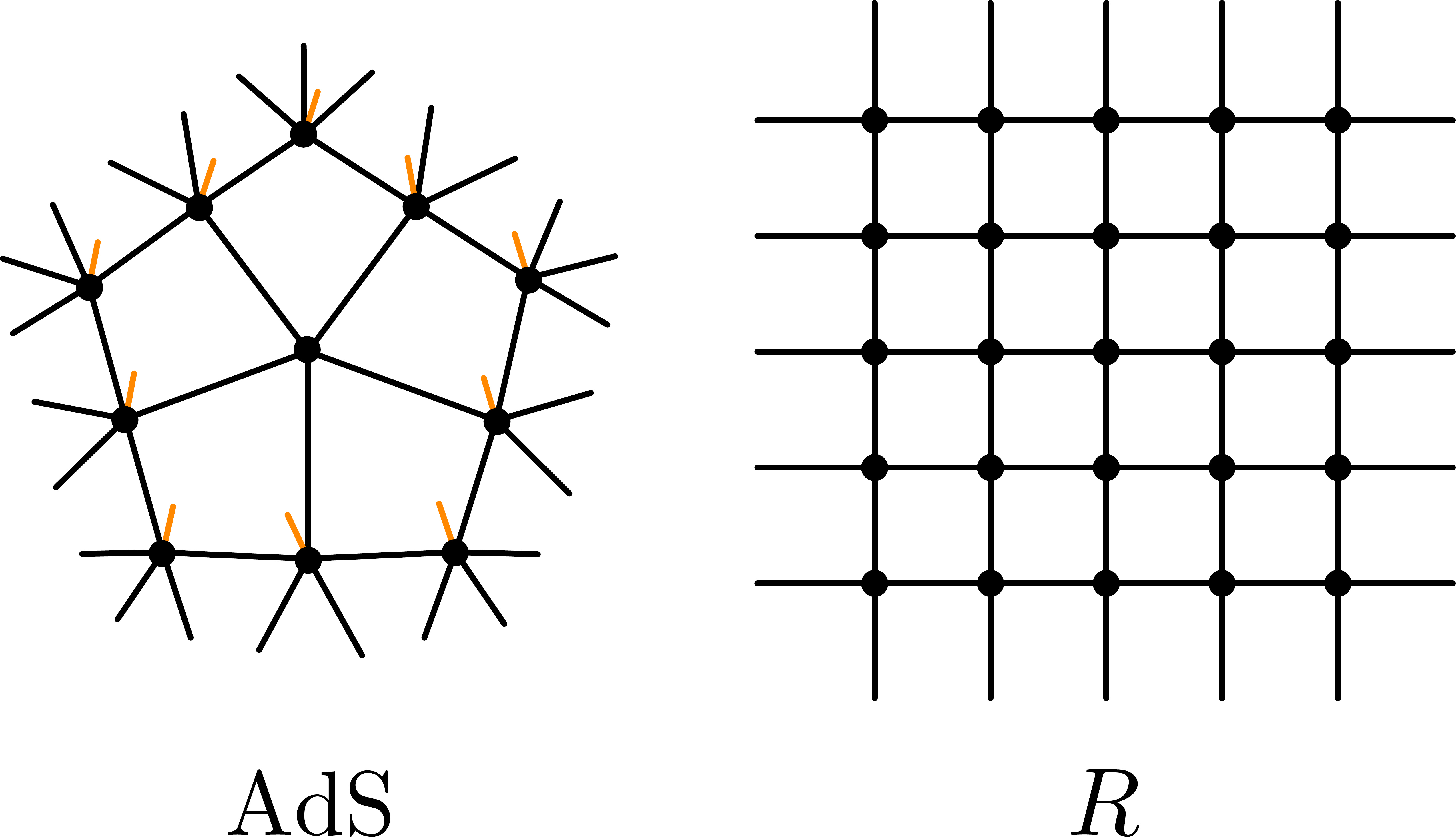}
    \caption{The tensor networks constructed from the tessellations of the AdS black hole and flat reservoir shown in Fig.~\ref{fig:AdS_R_Tess}}
    \label{fig:AdS_R_TNs}
\end{figure}

For example, the initial input bulk state at $t=0$ will consist of $m_0$ $\ell$ qudits in the AdS bulk. If $\cI \neq 0$, additional $R_\text{in}$ qudits representing future infalling degrees of freedom will need to be included as inputs to the reservoir tensor network. Applying the holographic tensor network (\ref{eq:AdS_R_PEPS}) then gives the fundamental state at $t=0$,
\begin{equation}
    |\Psi(t=0)\rangle = \Big( \langle\hat{\Psi}|_\ell \otimes \langle \hat{\Psi}|_{R_\text{in}} \Big) \Big( |\text{TN}\rangle_\text{AdS} \otimes |\text{TN}\rangle_R \Big)
\end{equation}
We note that since $\ell$ and $R_\text{in}$ are separable in the same tensor product as the AdS and $R$ tensor networks, the spacetimes remain disconnected. Fig.~\ref{fig:AdS_R_TNs} shows an example of these networks for the tessellations shown in Fig.~\ref{fig:AdS_R_Tess}.

\begin{figure}
    \centering
    \includegraphics[width=0.6\linewidth]{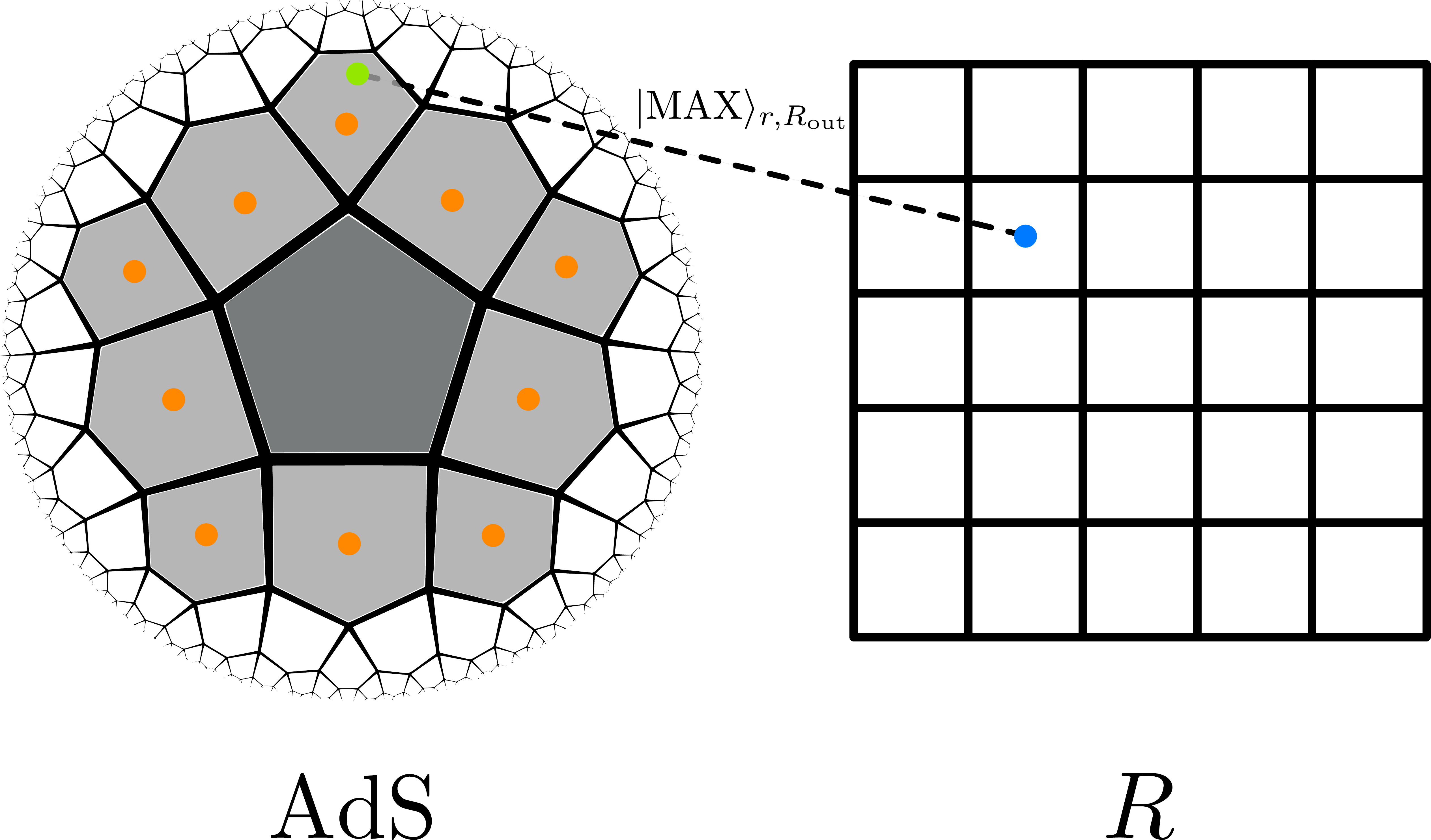}
    \caption{The interior and reservoir tessellations shown in Fig.~\ref{fig:AdS_R_Tess} evolved one fine-grained time step according to the effective dynamics, taking $\cI=0$ and $\cO=1$. A single maximally entangled Hawking pair has been added to the black hole interior and reservoir. A dashed line has been included to indicate that the \textcolor{green}{green} $r$ qudit and \textcolor{blue}{blue} $R_\text{out}$ qudit are maximally entangled.}
    \label{fig:HPair_Tess}
\end{figure}

Next, consider evolving the bulk state by one fine-grained time step. For simplicity, we will take $\cI = 0$ (no more qudits fall into the black hole after $t=0$) and $\cO = 1$ (only one Hawking pair is inserted in each time step). The new bulk state is then given by evolving the existing $\ell$ qudits with $\hat{U}_1$ and then inserting a single Hawking pair $|\text{MAX}\rangle_{r,R_\text{out}}$ into the bulk and reservoir. An example of the bulk and reservoir configurations of these degrees of freedom at $t=1$ is depicted in Fig.~\ref{fig:HPair_Tess}. The corresponding dual state can then be fed into the tensor network (\ref{eq:AdS_R_PEPS}) to obtain the state of black hole in the fundamental description at $t=1$,
\begin{equation} \label{eq:PEPS_t=1}
    |\Psi(t=1)\rangle =  \Big( \langle\text{MAX}|_{r,R_\text{out}} \otimes \langle\hat{\Psi}|_\ell \hat{U}_1^\dagger \Big) \Big( |\text{TN}'\rangle_\text{AdS} \otimes |\text{TN}'\rangle_R \Big),
\end{equation}
where we have used a prime on $|\text{TN}'\rangle$ as a reminder that the dimensions of each tensor state $|T_x\rangle$ need to be updated to accommodate the new degrees of freedom. In this way, (\ref{eq:AdS_R_PEPS}) provides the holographic map for which both the $\ell$ qudits and Hawking pairs are mapped to the fundamental description.

However, nothing prevents us from redefining the tensor network used to obtain (\ref{eq:PEPS_t=1}). Consider defining a new tensor network given by the base network (\ref{eq:AdS_R_PEPS}) and the new Hawking pair,
\begin{equation} \label{eq:newPEPS_t=1}
    |\text{TN}(t=1)\rangle = \langle\text{MAX}|_{r,R_\text{out}} \Big( \hat{U}_1^\dagger|\text{TN}'\rangle_\text{AdS} \otimes |\text{TN}'\rangle_R \Big),
\end{equation}
where $\hat{U}_1^\dagger$ acts to the right on the tensor network states $|T_x\rangle$ that receive $\ell$ as an input.
The fundamental state at $t=1$ can then be obtained by inserting the $\ell$ qudits into this new tensor network,
\begin{equation}
    |\Psi(t=1)\rangle = \langle\hat{\Psi}|_\ell\,|\text{TN}(t=1)\rangle.
\end{equation}
This gives a recursive process for constructing the holographic tensor network, where the tensor network $|\text{TN}(t)\rangle$ at time $t$ is defined to include the base network (\ref{eq:AdS_R_PEPS}) updated by $\hat{U}_t^\dagger$ and all Hawking pairs generated during the evolution.

\begin{figure}
    \centering
    \includegraphics[width=0.6\linewidth]{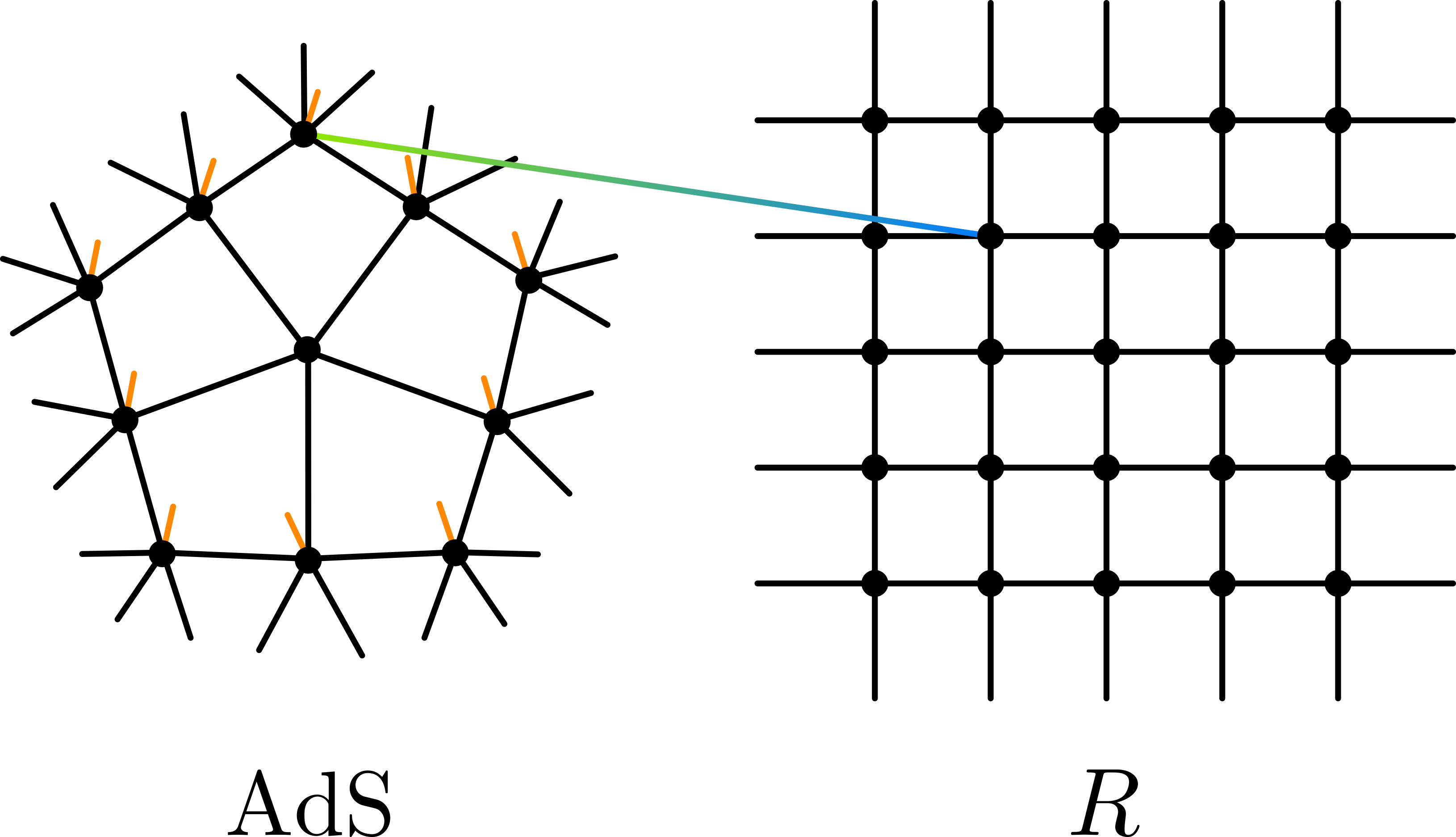}
    \caption{The tensor network constructed for the tessellation depicted in Fig.~\ref{fig:HPair_Tess} of the black hole interior and reservoir with one Hawking pair inserted. The maximally entangled state on $rR_\text{out}$ is interpreted as a new contraction between two tensors in the different networks, highlighted here as a \textcolor{green}{green}-\textcolor{blue}{blue} line.}
    \label{fig:HPair_TNs}
\end{figure}

Operationally, each Hawking pair in $|\text{TN}(t)\rangle$ is equivalent to the post-selection on $\langle\text{MAX}|$ used in (\ref{eq:contract}) to contract legs of neighboring tensors. Therefore, each new maximally entangled Hawking pair can be viewed as a new contraction between two tensors in the network. This new contraction is depicted in Fig.~\ref{fig:HPair_TNs} for $|\text{TN}(t=1)\rangle$ given in (\ref{eq:newPEPS_t=1}). Because the two components $r$ and $R_\text{out}$ of the Hawking pair live separately in the interior and reservoir (respectively), this new tensor network is no longer separable according to a tensor product between the original AdS and $R$ tensor networks. The two spacetimes must be connected. We therefore interpret the maximally entangled Hawking pairs as forming a wormhole between the two spacetimes!

\begin{figure}
    \centering
    \includegraphics[width=0.6\linewidth]{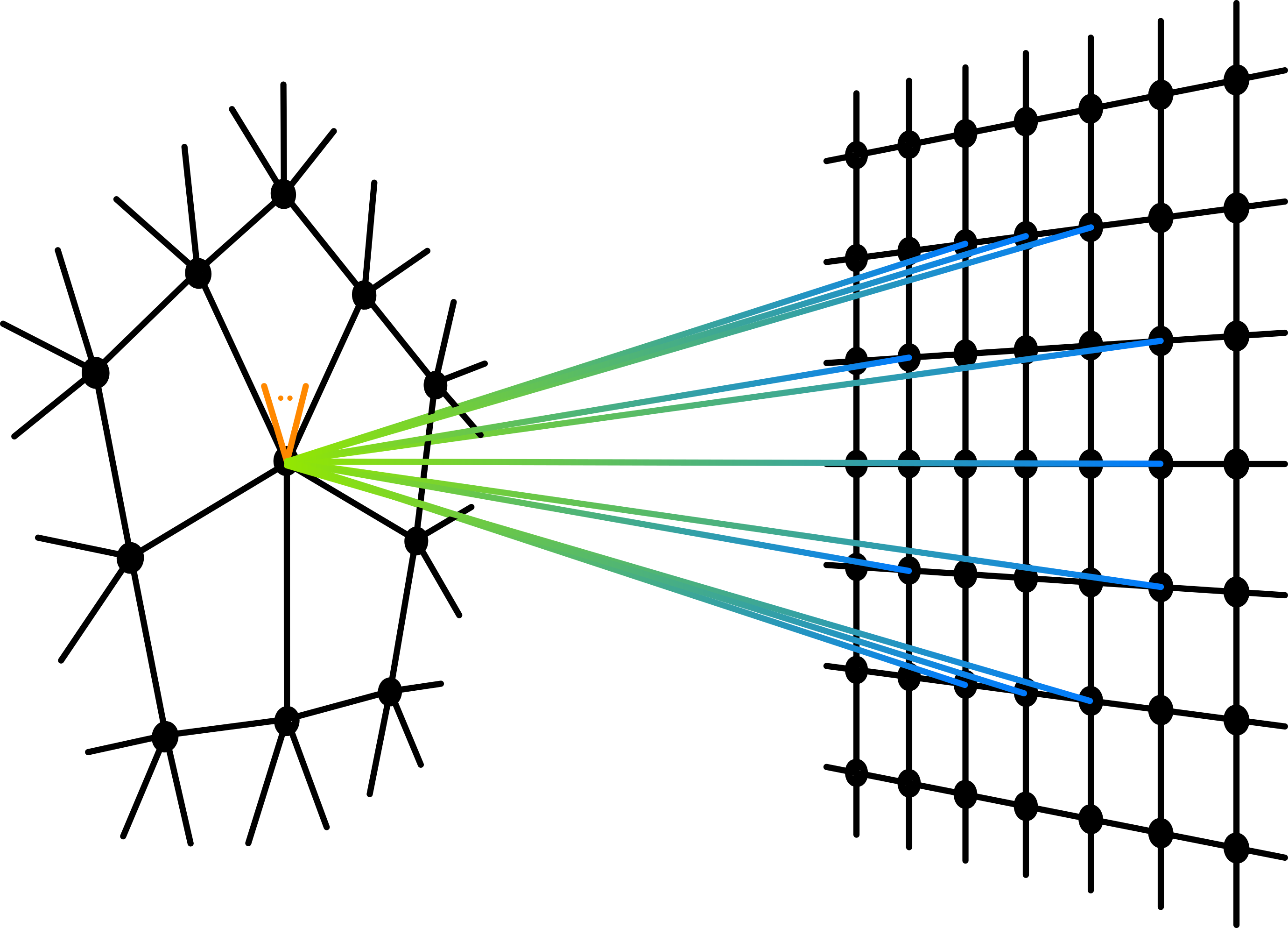}
    \caption{A depiction of a large wormhole (\textcolor{green}{green}-\textcolor{blue}{blue} lines) late in the evaporation of a black hole when the interior is comprised of only the $0^\text{th}$ row. A large number of new connections have been made between the AdS and reservoir tensor networks by the emission of a large amount of radiation over the course of the evaporation.}
    \label{fig:large_wormhole}
\end{figure}

As the black hole continues to evaporate, more Hawking pairs will be added to the network, creating more connections between the interior and the reservoir. In the coarse-grained dynamics of the tensor network, shedding one row can dramatically increase the number of these new connections. In this way, the wormhole will increase in ``width'' as the black hole evaporates. We note that it does not increase in length since each new connection should be thought of as having the same length. Fig.~\ref{fig:large_wormhole} depicts a large wormhole late in the evaporation of the black hole when the interior is only comprised of the $0^\text{th}$ row.

Had we instead chosen to make the reservoir a second AdS space and allowed the radiation $R_\text{out}$ to collapse into a second AdS black hole, these new connections would eventually lead to a two-sided black hole. We note that tensor networks for two-sided black holes have been considered in \cite{hartman_time_2013,brown_complexity_2016,brown_pythons_2020}; in all of these cases, the wormhole has more structure and is ``longer'' than our example, built out of its own tensors rather than just additional connections. This feature cannot be captured in our model, as we only describe evolution of states on top of an already existing tessellation. Dynamics of the tessellation itself could allow for the wormhole to grow in length. We leave exploration of these more sophisticated dynamics to future work.

\subsection{Dynamical QESs} 
Let us now return to studying the dynamics of QESs in evaporating black holes by applying the tools of Sec.~\ref{sec:QES} to these tensor networks for dynamically generated states. From our above understanding of maximally entangled legs as new contractions between the bulk and reservoir tensor networks, we can understand the contributions from unpaired bulk $r$ legs to $S(EW)$ as a contribution from the area of the wormhole connecting the bulk to the reservoir. This is reminiscent of the ``expandable space blimp'' geometry introduced in \cite{brown_pythons_2020}; see Fig.~14 of that work for a schematic of the sweeping process that illustrates this counting.

Our goal will be to reproduce the QES dynamics established in \cite{penington_entanglement_2020,almheiri_entropy_2019} for an evaporating black hole. As mentioned in Sec.~\ref{sec:op_recon}, we expect to find two candidate QESs: one zero-area surface in the $0^\text{th}$ row and one non-trivial surface in the outermost row behind $\Lambda_\text{BH}$. We expect the latter QES to form in the initial moments of the black hole's evaporation, diagnosing its appearance by $\delta S_\text{gen} \leq 0$ in the outermost row. Early in the black hole's evaporation, the zero-area surface should be the global minimum such that $\Delta S_\text{gen} > 0$. We expect a phase transition to occur at the Page time such that the non-trivial QES becomes minimal as indicated by $\Delta S_\text{gen} \leq 0$. The entire inner code must have become fully non-isometric by this time, as described in Sec.~\ref{sec:QES}.

We can choose to study the change in $S_\text{gen}$ of either $B$ or $R_\text{out}$ to diagnose the behavior of the non-trivial QES. In \cite{dewolfe_bulk_2024}, the generalized entropy of $R_\text{out}$ was used to find the Page time as
\begin{equation} \label{eq:PageTime}
    t_\text{Page} = \frac{n_0 + S(R_\text{in})}{2\cO - \cI}.
\end{equation}
We will do the same here, using the local change $\delta S_{\text{gen},R_\text{out}}(N)$ and global change $\Delta S_{\text{gen},R_\text{out}}(N)$ to study the non-trivial QES. Furthermore, we will make the simplifying assumption that $\hat{U}_t = \mathbb{1}$ so that the interior degrees of freedom can move freely but will not interact with each other. Non-trivial dynamics $\hat{U}_t \neq \mathbb{1}$ would allow the entropy of each bulk qudit to spread out amongst other bulk qudits, complicating the following analysis. We will also use this assumption to restrict the $r$ qudits to the outermost row (since they are moving radially outward) while the $\ell$ qudits move deeper into the interior.

Let us begin with an interior tensor network of initial size $N_0$ and let one row evaporate; now, the outermost row is $N_0 - 1$. For simplicity, let us assume that $\Delta T(N_0)$ is large enough so that the $\ell$ qudits have had enough time to move deeper into the interior so that none are in row $N_0-1$. The local change in $S_{\text{gen},R_\text{out}}$ across the outermost row is then
\begin{equation}
    \delta S_{\text{gen},R_\text{out}} (N_0 - 1) = q_B(N_0-1) - q_B(N_0-2) - \frac{\cO}{\cO-\cI} \Big( q_B(N_0) - q_B(N_0-1) \Big),
\end{equation}
where we have used (\ref{eq:bulkDOFs_perRow}) to replace the $q_r(N)$ term in (\ref{eq:local_Srout}). We can make use of the large $N_0$ limit of the fractional difference between rows given by (\ref{eq:largeN}) by dividing by $q_B(N_0-1)$, a strictly positive quantity:
\begin{equation} \label{eq:largeN_localChange}
    \lim_{N_0\rightarrow\infty} \frac{ \delta S_{\text{gen},R_\text{out}} (N_0 - 1)}{q_B(N_0-1)} = - \frac{k(n-2) - 2n}{2(\cO-\cI)} \Bigg[ (2\cO - \cI) + \cI \sqrt{\frac{(n-2)(k-2)}{k(n-2) - 2n}} \Bigg].
\end{equation}
We see that the local change in $S_{\text{gen},R_\text{out}}$ across the outermost row is strictly negative so long as $\cO > \cI$ (which is guaranteed for an evaporating black hole) and
\begin{equation}
    k \geq \frac{2n}{n-2}.
\end{equation}
We recognize this second condition as the requirement for the tensor network to be flat (equality) or negatively curved. Therefore, a non-trivial QES is guaranteed to form in the outermost row as soon as one row has evaporated in either flat or negatively curved interior tensor networks. This confirms the first of our two expectations for QES dynamics.

Next, we use the global change to determine if this non-trivial QES is the global minimum. Using (\ref{eq:global_Rout}), the global change in $S_{\text{gen},R_\text{out}}$ is
\begin{equation}
    \Delta S_{\text{gen},R_\text{out}} ( N_0-1) = q_B(N_0 - 1) - \frac{\cO - s\cI}{\cO - \cI} \Big( q_B(N_0) - q_B(N_0-1) \Big).
\end{equation}
Again dividing by $q_B(N_0 - 1)$ and taking the large $N_0$ limit, we find
\begin{equation}
    \lim_{N_0\rightarrow\infty} \frac{\Delta S_{\text{gen},R_\text{out}} (N_0-1)}{q_B(N_0-1)} = 1 - 2 \frac{\cO - s\cI}{\cO-\cI} \left( -1 + \sqrt{\frac{(n-2)(k-2)}{k(n-2) - 2n}} \right)^{-1}.
\end{equation}
The non-trivial QES will transition to the global minimum when this global change is $\leq0$; this requires
\begin{equation} \label{eq:globalmin_cond}
    n \geq 2 + \frac{4x^2}{\tilde{k}(1-x^2)},
\end{equation}
where we have defined
\begin{equation}
    \tilde{k} \equiv k - \frac{2n}{n-2}\in\mathbb{Z} \qquad \text{and} \qquad 
    x \equiv \frac{\cO - \cI}{3\cO - (2s + 1) \cI}.
\end{equation}
We note that $\tilde{k}$ describes how negatively curved the interior tensor network is: $\tilde{k} = 0$ corresponds to a flat network, while $\tilde{k}\geq1$ describes a negatively curved network. The variable $x$ captures information about the infalling matter and outgoing radiation. We note that $x$ is always positive when $\cO > \cI$, as is the case for an evaporating black hole. Additionally requiring $0\leq s\leq 1$, the second term of (\ref{eq:globalmin_cond}) is bounded by
\begin{equation}
    0 < \frac{4x^2}{1-x^2} \leq \frac{1}{2}.
\end{equation}
Because of this, the greatest lower bound on $n$ given by (\ref{eq:globalmin_cond}) for a hyperbolic tensor network is $n\geq5/2$. Since polygons require $n\geq3$, the condition (\ref{eq:globalmin_cond}) is then satisfied for all hyperbolic networks, indicating that the non-trivial QES that formed after one row evaporated is automatically the global minimum. The Page time must have occurred before a single row has evaporated!

We note that for flat tensor networks ($\tilde{k} = 0$), the right hand side of (\ref{eq:globalmin_cond}) is infinite, and therefore no finite $n$ can satisfy the inequality. This indicates that the non-trivial QES that formed in row $N_0 - 1$ is not yet the global minimum, and the Page time has not yet been reached after one row has evaporated. We can find the Page time for flat networks by determining the number of rows $t_\text{row}$ that must evaporate before the global change $\Delta S_{\text{gen},R_\text{out}} (N_0 - t_\text{row})$ becomes negative. Using $S(R_\text{in})$ for the total entropy of all internal $\ell$ degrees of freedom, setting $\Delta S_{\text{gen},R_\text{out}} (N_0 - t_\text{row})$ to zero gives
\begin{equation} \label{eq:trow_flat}
    t_\text{row} = \frac{\cO - \cI}{2\cO - \cI} \left( N_0 + \frac{1}{2} + \frac{S(R_\text{in})}{2n} \right).
\end{equation}
Since $0 \leq (\cO - \cI)/(2\cO - \cI) \leq 1/2$, this is typically larger than just one row. Therefore, there is a distinction between the formation of the non-trivial QES and its phase transition at the Page time in flat tensor networks. The time $t$ elapsed after $t_\text{row}$ rows have evaporated is found by using (\ref{eq:dT}),
\begin{equation}
    t = \sum_{j=0}^{t_\text{row}-1} \Delta T(N-j).
\end{equation}
For flat networks, this simplifies to
\begin{equation}
    t = \frac{2n}{\cO - \cI} t_\text{row}, \qquad \text{flat network.}
\end{equation}
Substituting this into (\ref{eq:trow_flat}), we find the Page time
\begin{equation} \label{eq:PageTime_TN}
    t_\text{Page} = \frac{q_B(N_0) + S(R_\text{in})}{2\cO - \cI},
\end{equation}
where we used 
\begin{equation}
    q_B(N_0) = n(1 + 2N_0), \qquad \text{flat network},
\end{equation}
to replace $N_0$ with $q_B(N_0)$. Recognizing that $q_B(N_0) = n_0$, this exactly agrees with the Page time (\ref{eq:PageTime}) found in \cite{dewolfe_bulk_2024}.

In summary, we have indeed found that a non-trivial QES forms in the outermost row of the interior tensor network as soon as one row has evaporated, as indicated by (\ref{eq:largeN_localChange}). Unfortunately, this non-trivial QES is automatically the global minimum for hyperbolic tensor networks, so we cannot observe the phase transition of the QES. This is consistent with the over $50\%$ fractional loss of $q_B(N)$ in hyperbolic tessellations discussed at the end of Sec.~\ref{sec:noniso_structure}. While studying flat networks is not the goal of the present work, we do find that the phase transition is resolvable by the coarse-grained dynamics for flat tensor networks. Furthermore, the corresponding Page time exactly reproduces that found in \cite{dewolfe_bulk_2024}. This suggests that the phase transition does occur for the hyperbolic networks, but the coarse-grained time evolution is simply too coarse to resolve it, $\Delta T(N_0) > t_\text{Page}$. 

\subsection{Interior reconstruction for dynamically generated states}

We now return to the discussion of interior state and operator reconstruction from Sec.~\ref{sec:op_recon}. There, we described how interior states and operators would be encoded in $B$ by the interior tensor network long before the Page time, but had yet to demonstrate reconstructability in the radiation $R_\text{out}$ long after the Page time. We take up this question here using tensor networks for dynamically generated states described in Sec.~\ref{sec:wormhole}. As in Sec.~\ref{sec:wormhole}, we interpret the maximally entangled $r$ legs as forming new contractions between the interior and reservoir tensor networks, building the tensor network analog of a wormhole connecting the two spacetimes.

Long before the Page time, only a small amount of radiation has been generated during the short period of evaporation, creating only a few new connections to form a small wormhole. Regardless, the interior and reservoir tensor networks are still connected, and an initial state $|\hat{\Psi}\rangle_{R_\text{in}}$ inserted into $|\text{TN}(t)\rangle$ will be mapped to the boundaries of both spacetimes. During these initial stages when $|B|$ is still much larger than both  $|\ell|$ and $|r|$, we expect that only a small amount of the information contained in $|\hat{\Psi}\rangle_{R_\text{in}}$ will leak to the reservoir, leading to a reconstruction of the initial state in $B$ to some high precision. Similarly, when pushing operators acting on $\ell r$ through the network, the resulting reconstruction in the fundamental description will have highest precision on the boundary of the bulk AdS due to the lack of wormhole connections. Therefore, so long as a condition like (\ref{eq:approx_isom_cond}) is satisfied, an approximately isometric encoding of both states and operators in $B$ by the interior tensor network is still possible. The pre-Page time results of Sec.~\ref{sec:op_recon} then still hold true.

Long after the Page time, there will be significantly more radiation forming a large wormhole connecting the interior to the reservoir. There will also be much fewer fundamental degrees of freedom $B$; in fact, the total number of wormhole connections given by $\log_D|r|$ will have surpassed the number of $B$ legs when
\begin{equation}
    t > \frac{n_0}{2\cO - \cI},
\end{equation}
which is very close to the Page time given in (\ref{eq:PageTime}) and (\ref{eq:PageTime_TN}). Thus while the AdS portion of $|\text{TN}(t)\rangle$ acts as a strong non-isometry on an initial state $|\hat{\Psi}\rangle_{R_\text{in}}$ long after the Page time, the large wormhole will serve as an isometric escape for the $\ell$ degrees of freedom. Thus the interior state will be reconstructable (to a high precision) in the reservoir long after the Page time. 

Similarly, the large wormhole could allow for an interior operator to be pushed through the new connections, giving a reconstruction (to some precision) in the reservoir long after the Page time. The process for doing so bears more investigation since small interior operators (possibly at early times when there are fewer qudits per cell) and large interior operators (possibly at late times when a large number of qudits have been forced to a small number of central cells) may need to be pushed through the wormhole in different ways. We will leave the investigation of this interesting topic to future work.

Thus, we see that this model appropriately encodes interior states and operators in $B$ long before the Page time, while a wormhole formed by the insertion of maximally entangled Hawking pairs provides a mechanism for encoding in the reservoir $R_\text{out}$ long after. This reproduces expectations from both entanglement wedge reconstruction and PHEVA's non-isometric maps, while providing geometric intuition through the formation of a wormhole in the tensor network.

\section{Adding locality to the BFP map} \label{sec:BFP}

In \cite{akers_black_2022}, PHEVA additionally proposed a dynamical holographic map as a special construction of their generic map. This map replaced the Haar random unitary $U_H$ in (\ref{eq:V_PHEVA}) with the dynamics of the fundamental description,
\begin{equation}
    U \equiv U_t U_{t-1} \dots U_0,
\end{equation}
where (as in Sec.~\ref{sec:dynamics}) each $U_t$ is drawn from the Haar measure. Furthermore, the post-selection $\langle\phi|_P$ in (\ref{eq:V_PHEVA}) was was replaced with post-selection on the Hawking pairs $rR_\text{out}$ in the maximally entangled state $\langle \text{MAX}|_{r,R_\text{out}}$. Thus their dynamical holographic map can be expressed as 
\begin{equation} \label{eq:PHEVA_dyn}
    V = |r|\langle\text{MAX}|_{r,R_\text{out}} U |\psi\rangle_f,
\end{equation}
where $|r| = \sqrt{|P|}$ is included to preserve normalization. Importantly, this model assumed that the effective dynamics,
\begin{equation}
    \hat{U} \equiv \hat{U}_t \hat{U}_{t-1} \dots \hat{U}_0
\end{equation}
were trivial ($\hat{U}=\mathbb{1}$) so that the effective degrees of freedom could be fed directly into the fundamental dynamics.

Recent work has sought to generalize this dynamical map to include non-trivial effective dynamics. First, Kim and Preskill included non-trivial dynamics in the input to the dynamical map without changing the map itself \cite{kim_complementarity_2023}. They found small violations in the unitarity of black hole evaporation; because these violations were exponentially suppressed in the black hole's entropy, they concluded that small changes to the effective description were needed to restore unitarity. Furthermore, they discussed possible exponential increases in computational complexity in violation of the quantum extended Church-Turing thesis \cite{aaronson_quantum_2005,deutsch_quantum_1997,susskind_horizons_2020} and superluminal signaling \cite{bao_grover_2016}. 

\begin{figure}
    \centering
    \input{transformations}
    \caption{Circuit diagrams for the BF (left) and BFP (right) maps. The middle circuit diagram gives the transformation used to equate the two maps on the dynamically generated subspace of the effective description. Figure reproduced from \cite{dewolfe_bulk_2024}.}
    \label{fig:transformations}
\end{figure}
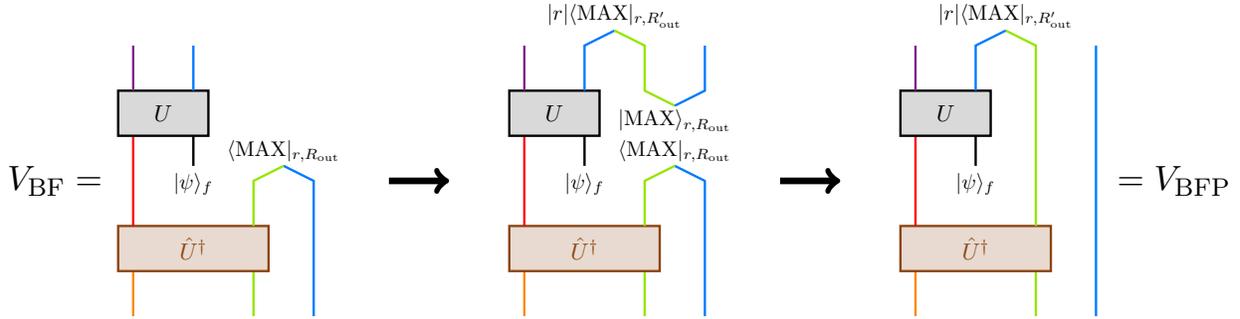

These problems were remedied in \cite{dewolfe_non-isometric_2023,dewolfe_bulk_2024} by incorporating the non-trivial effective dynamics into the holographic map itself. The resulting ``backwards-forwards'' (BF) map begins by backwards evolving the effective degrees of freedom $\ell r$ using $\hat{U}^\dagger$ to bring them back out of the black hole. Hawking radiation annihilates and returns to the vacuum during the backwards evolution, represented as post-selection on $\langle\text{MAX}|_{r,R_\text{out}}$. Once all internal degrees of freedom have either disappeared or are back in $R_\text{in}$, forwards fundamental dynamics can be performed to bring the black hole into the fundamental description. All together, the BF map is given as
\begin{equation}
    V_\text{BF} = U|\psi\rangle_f \langle\text{MAX}|_{r,R_\text{out}} \hat{U}^\dagger
\end{equation}
and is represented as a circuit diagram in the left of Fig.~\ref{fig:transformations}.

The ``backwards-forwards-post-selection'' (BFP) map was also proposed in \cite{dewolfe_non-isometric_2023} and shown to be equivalent to the BF map on the dynamically generated subspace,
\begin{equation}
    V_\text{BF} \hat{P}_t = V_\text{BFP} \hat{P}_t,
\end{equation}
where $\hat{P}_t$ is the projection onto the dynamically generated subspace of the effective description. The BFP map delays post-selection until the end of the map, and post-selects on the radiation output of the fundamental dynamics $R'_\text{out}$ instead of the reservoir. During the backwards effective dynamics, Hawking pairs $rR_\text{out}$ are ``frozen out'' at some high energy to prevent the formation of past singularities. All together, the BFP map is given as
\begin{equation}
    V_\text{BFP} = |r| \langle\text{MAX}|_{r,R'_\text{out}} U |\psi\rangle_f \hat{U}^\dagger
\end{equation}
and is represented as a circuit diagram in the right of Fig.~\ref{fig:transformations}. The middle diagram in Fig.~\ref{fig:transformations} shows the transformation that equates the BF and BFP maps on the dynamically generated subspace.

The BFP map has several important features worthy of noting. First, it reduces to PHEVA's dynamical map (\ref{eq:PHEVA_dyn}) in the limit of trivial effective dynamics $\hat{U} = \mathbb{1}$. Second, it was shown in \cite{dewolfe_bulk_2024} to satisfy all requirements set by PHEVA in \cite{akers_black_2022} for a well-behaved holographic map. In particular, it acts as an exact isometry on dynamically generated states, acts isometrically on average (with exponentially suppressed deviations) on any generic state of the effective Hilbert space, provides a state-dependent reconstruction of interior operators to exponential accuracy, and reproduces the Page curve.

Given the success of the BFP map in generalizing the dynamical map proposed by PHEVA, we would like to add a notion of locality to it similar to the interior tensor networks described above. This can also be done with hyperbolic tessellations given the dynamics of both descriptions from Sec.~\ref{sec:dynamics}. We first construct the BF map and then use the transformations depicted in Fig.~\ref{fig:transformations} to obtain the BFP map.

\begin{figure}
    \centering
    \includegraphics[width=\linewidth]{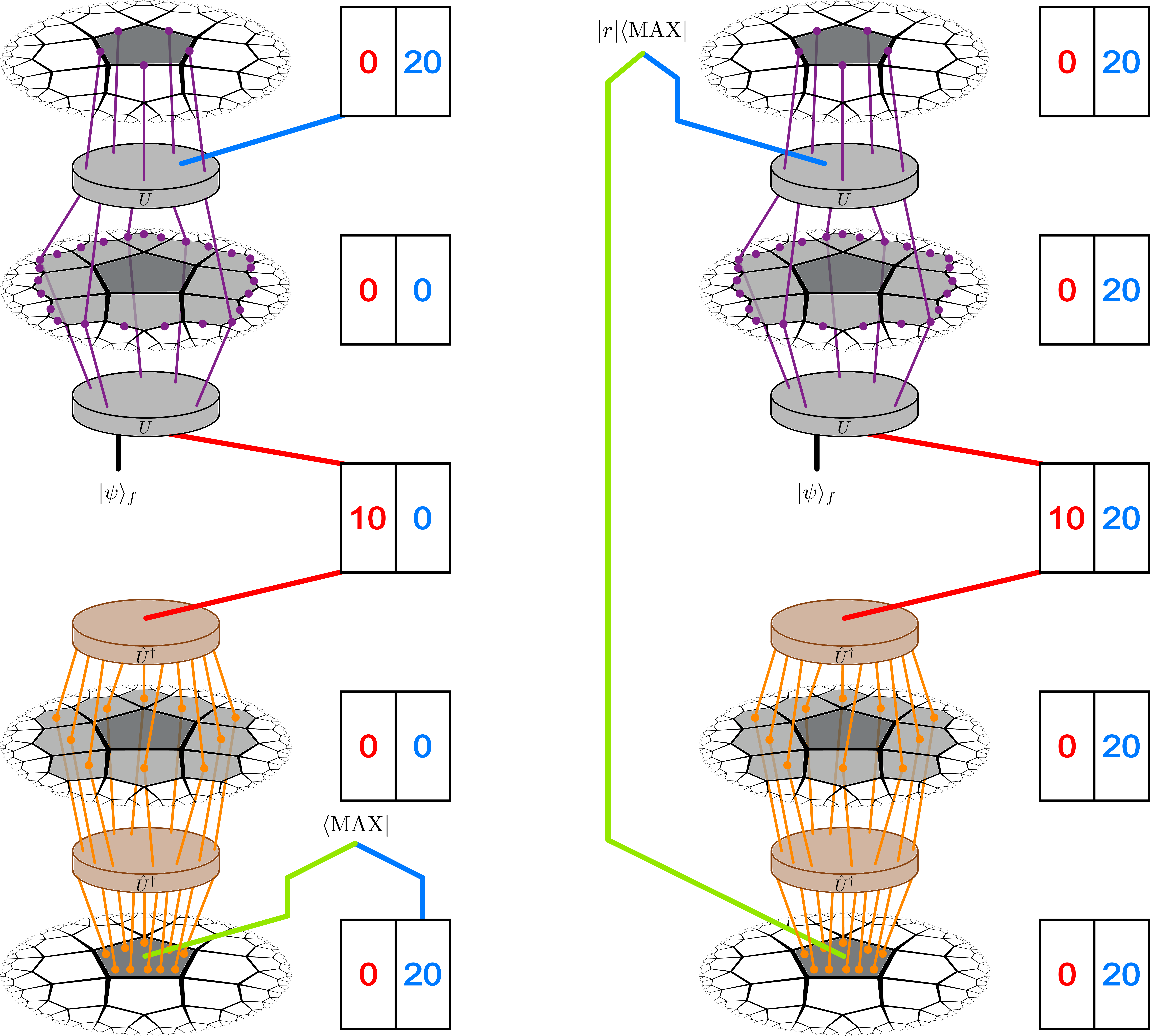}
    \caption{Circuit diagrams depicting the BF (left) and BFP (right) maps for a black hole initially containing two rows of cells in a $\{5,4\}$ tessellation. Qudits colored by type according to Fig.~\ref{fig:colors}. Thin lines denote a single qudit; thick lines denote all qudits of that type. Note that not all purple lines are drawn to avoid an overcrowding of lines.}
    \label{fig:BFandBFP_N2}
\end{figure}

As before, the BF map begins by running the effective dynamics described in Sec.~\ref{sec:dynamics} backwards. Post-selection on $\langle\text{MAX}|_{r,R_\text{out}}$ is performed on $r$ qudits in the tessellation and $R_\text{out}$ qudits in the reservoir to remove them as Hawking pairs annihilate back to the vacuum during the backwards evolution. Once all effective dynamics are undone, all qudits will have returned to the reservoir as $R_\text{in}$ qudits. The forwards fundamental dynamics of Sec.~\ref{sec:dynamics} can then be applied, bringing the tessellation and associated degrees of freedom into the fundamental description. The map finishes after an equal amount of time evolution has been performed in the fundamental description. The resulting BF map is depicted as a circuit diagram in the left panel of Fig.~\ref{fig:BFandBFP_N2} for the case of a $\{5,4\}$ tessellation with $m_0 = 10$, $N_0 = 2$, and $\cI=0$.

The BFP map is then obtained by the transformations described in Fig.~\ref{fig:transformations}. Now instead of post-selecting on $\langle\text{MAX}|_{r,R_\text{out}}$ during the backwards effective evolution, the $r$ and $R_\text{out}$ degrees of freedom are ``frozen out'' and stored in a quantum memory as detailed in \cite{dewolfe_bulk_2024}. Post-selection on $\langle\text{MAX}|_{r,R'_\text{out}}$ is only performed after the fundamental evolution, where $R'_\text{out}$ denotes the radiation outputted by the fundamental dynamics. The resulting BFP map is depicted as a circuit diagram in the right panel of Fig.~\ref{fig:BFandBFP_N2} for the case of a $\{5,4\}$ tessellation with $m_0 = 10$, $N_0 = 2$, and $\cI=0$.

So long as the local unitaries $\hat{U}_t$ from Sec.~\ref{sec:dynamics} can be expressed as some $k$-design, the conclusions of \cite{dewolfe_bulk_2024} on the behavior of the BFP map then apply here. On dynamically generated states (those prepared by the dynamics shown in the top sequence of Fig.~\ref{fig:dynamics_N2}), the BFP map with locality will act isometrically without need for averaging. On generic states, it will act isometrically on average (with exponentially small variations) and provide a state-dependent reconstruction of interior operators (to exponential precision). This reconstruction implies bounds for entanglement wedge reconstruction and reproduces the Page curve. All of these behaviors are expected of a holographic map from \cite{akers_black_2022}.

Since the BFP map can be made local using hyperbolic tessellations in the same way as the interior tensor networks described above, we take this construction of the BFP map as a special case of the interior tensor networks. The fact that it behaves well on generic states gives further evidence that the more generic tensor network constructions of Sec.~\ref{sec:TN} can be constructed to satisfy the same properties.

\section{Conclusion} \label{sec:conc}

In this work, we have used hyperbolic tessellations of AdS black hole interiors to assign a notion of locality (on an AdS length scale) to qudits describing interior degrees of freedom. From these tessellations and their associated bulk degrees of freedom, we constructed tensor networks to give a generalization of PHEVA's non-isometric codes to include a more refined notion of locality. Non-isometric tensors were incorporated into the network by allowing for an arbitrary number of inputs from bulk degrees of freedom. These interior tensor networks define an ``inner code'' that maps the effective degrees of freedom $\ell$ and $r$ defined in \cite{akers_black_2022} to fundamental black hole degrees of freedom $B$ living on the horizon. Exterior tensor networks then isometrically encode $B$ in the CFT degrees of freedom living on the boundary of the bulk spacetime, providing the ``outer code'' that completes the holographic map. So long as the unitaries used to construct the interior tensor network combine to act as a pseudorandom unitary or unitary $k$-design on the entire space of states, this holographic map should act similarly to PHEVA's non-isometric map on generic states of the black hole interior.

The locality added by the interior tensor networks gave us the ability to study the structure of non-isometries in the inner code. Most interestingly, we found a connection between non-isometric rows and QESs in Sec.~\ref{sec:TN}: the presence of a QES in a row of the tessellation implies the existence of non-isometric tensors in that row of the tensor network. The behavior of the entire inner code is greatly affected by the position and strength of these non-isometric rows. In the case of an evaporating black hole, we expect that non-isometries associated with a single non-trivial QES in the outermost row will only be weakly non-isometric (as defined in Sec.~\ref{sec:noniso_structure}) long before the Page time. Thus the inner code can be described by an approximate isometry satisfying (\ref{eq:approx_isom}), allowing for interior states and operators to be encoded to some high precision in the boundary CFT.
As the non-trivial QES transitions to the global minimum at the Page time, we expect that the non-isometries will grow in strength until an approximately isometric encoding in the CFT is no longer possible.

To investigate further, we defined a limited notion of time evolution for both descriptions. Though time evolution in tensor networks is an open problem, we were able to implement coarse-grained dynamics by assigning degrees of freedom to cells within the tessellation based on the models of evolution described in \cite{akers_black_2022,dewolfe_non-isometric_2023,dewolfe_bulk_2024}. We found that tensor networks for bulk states prepared by the effective dynamics form wormholes from the insertion of Hawking pairs. These wormholes stretch between the black hole interior and the reservoir storing the radiation and provide a mechanism for encoding interior states and operators in the radiation long after the Page time.

Furthermore, we used these dynamics to study QESs in an evaporating black hole. We were able to observe a non-trivial QES forming in the outermost row of the tessellation after the black hole evaporated by one coarse-grained time step. However, we were not able to observe the QES phase transition at the Page time for any hyperbolic tessellation; the non-trivial QES was automatically the global minimum in all cases. Interestingly, we were able to see the QES phase transition and compute the Page time for flat tessellations. While flat tessellations are not the focus of this work, we take this to indicate that a hyperbolic tessellation with more fine-grained dynamics could resolve the Page time. 

We leave resolving the Page time in hyperbolic interior tensor networks to future work. There are a few possible avenues for exploring this. First, increasing the resolution of the network itself by adding sub-AdS length scale locality would allow for smaller changes in the black hole's size as the black hole sheds one row. This would in turn increase the resolution of the coarse-grained dynamics described in Sec.~\ref{sec:dynamics}, possibly resolving the Page time. Second, more sophisticated tensor network constructions incorporating gauge degrees of freedom \cite{donnelly_living_2017,qi_emergent_2022,dolev_gauging_2022,dong_holographic_2024,akers_background_2024,akers_multipartite_2024} could lead to a better understanding of time dynamics for tensor networks, possibly providing a way to resolve the Page time. 

As a final comment, we note that other works have considered other aspects of non-isometry and locality in holography. The work of \cite{cao_overlapping_2024} uses non-isometric maps to construct ``overlapping qubits'' that can be used to model a local effective field theory with a system that contains many fewer degrees of freedom. Here we have focused on locality for PHEVA-like holographic maps, and it would be interesting to further compare these results. We are hopeful that further improvements in these non-isometric codes for holography and black holes will continue to push our understanding of quantum gravity and black hole information.

\section*{Acknowledgements}

We are grateful to Chris Akers for helpful discussions, and to Brian Swingle and Wissam Chemissany for discussions on related work. The images of tessellations used in figures throughout this paper were generated using the \href{https://www.geometrygames.org/KaleidoTile/index.html.en}{KaleidoTile} program. The authors are supported by the Department of Energy under grants DE-SC0010005 and DE-SC0020360.

\appendix

\section{Tessellation counting} \label{app:counting}

\begin{figure}
    \centering
    \includegraphics[width=0.5\linewidth]{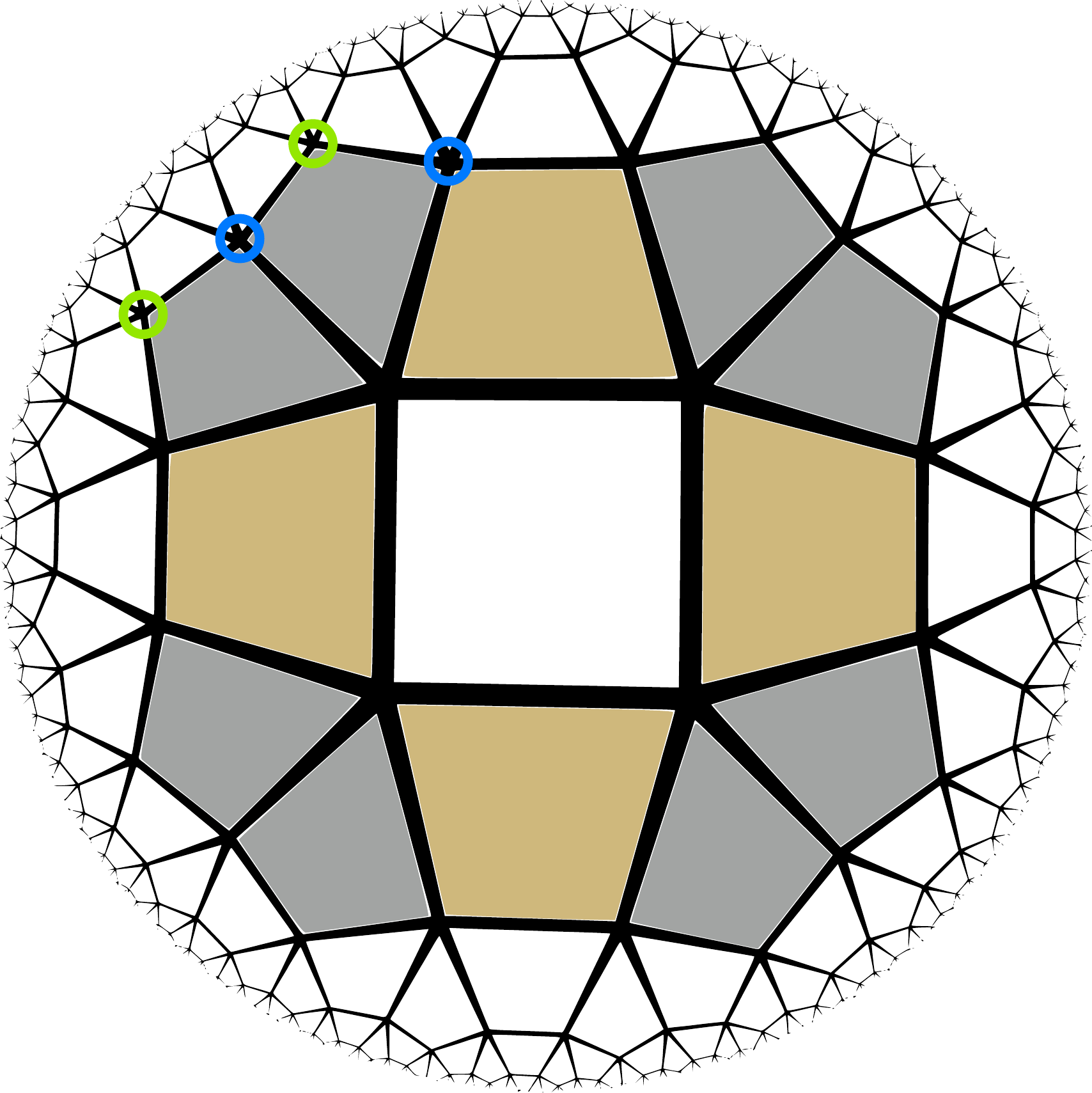}
    \caption{A depiction of a $\{4,5\}$ tessellation. The cells in row 1 are colored according to their type used for counting: ``edge cells'' are in gold, and ``vertex cells'' are in gray. Two types of vertices are identified on the outside of row 1 as well: vertices circled in blue are shared between two cells in row 1, and vertices circled in green are unshared.}
    \label{fig:counting}
\end{figure}

Here we provide a discussion of the counting done by equation (\ref{eq:facevec}) for the number of cells in each row of a tessellation. As before, $f_e$ counts the number of cells in row $N$ attached by an edge to row $N-1$, while $f_v$ counts the number of cells in row $N$ attached by a vertex to row $N-1$. We will refer to each type of cell by the shorthands ``edge cell'' and ``vertex cell'', respectively. Fig.~\ref{fig:counting} provides a depiction of a $\{4,5\}$ tessellation that will be useful to reference throughout this discussion.\footnote{We use a $\{4,5\}$ tessellation as an example here instead of the $\{5,4\}$ example used throughout the main body of this work because it will demonstrate features of the counting that are not visible in a $k=4$ tessellation.}

Since we have chosen to draw each tessellation such that there is a single cell at the center, row 0 will always have one cell. Counting then begins in row 1. Since the cell in row 0 has $n$ edges, there will be $n$ edge cells in row 1. These are colored gold in Fig.~\ref{fig:counting}. The central cell also has $n$ vertices, and each of these will contribute $k-3$ vertex cells to row 1 since there are $k$ cells around a vertex and three are already taken up by the central cell and two edge cells. These are colored gray in Fig.~\ref{fig:counting}. Thus for $N=1$ we obtain $(f_e\,\,\,\, f_v) = (n\,\,\,\, n(k-3))$; this gives the initial vector on the right hand side of (\ref{eq:facevec}).

For each subsequent row, all cells in row $N-1$ contribute both edge and vertex cells to row $N$. The contribution to row $N$ from each cell in row $N-1$ is linear and fixed by $n$ and $k$; determining the vector $(f_e\,\,\,\, f_v)$ for row $N$ can then be done by multiplying $(f_e\,\,\,\, f_v)$ for row $N-1$ by a fixed $2\times2$ matrix $M$ completely determined by the Schl\"{a}fli symbol $\{n,k\}$ of the tessellation. Tracing this all the way down to row 1, we obtain a schematic representation of (\ref{eq:facevec}),
\begin{equation}
    \begin{pmatrix}
        f_e \\
        f_v
    \end{pmatrix}
    = M^{N-1}
    \begin{pmatrix}
        n \\
        n(k-3)
    \end{pmatrix}.
\end{equation}

The top row of $M$ gives the contribution of each cell in row $N-1$ to edge cells in row $N$. Each cell in row $N-1$ contributes one edge cell to row $N$ for each edge not already shared by another cell in row $N-1$ or $N-2$. Since edge cells in row $N-1$ have three edges occupied (one by row $N-2$ and two by neighbors in row $N-1$) they contribute $n-3$ edge cells to row $N$. Similarly, vertex cells in row $N-1$ have two edges occupied (both by neighboring cells in row $N-1$) so they contribute $n-2$ edge cells to row $N$. The top row of $M$ is then given by
\begin{equation}
    M = 
    \begin{pmatrix}
        n-3 &   n-2 \\
        ?   &   ?
    \end{pmatrix}.
\end{equation}

The bottom row of $M$ gives the contribution of each cell in row $N-1$ to vertex cells in row $N$. Each cell in row $N-1$ contributes some number of vertex cells (depending on $k$) for each of their vertices on the outside of row $N-1$. There are two types of vertices: vertices shared by two neighboring cells in row $N-1$, and unshared vertices belonging to only one cell. These two types are identified by blue and green circles (respectively) in Fig.~\ref{fig:counting}. Consider the shared vertices first. There is one shared vertex for every cell in row $N-1$; each contributes $k-4$ new vertex cells, since two cells around the vertex are taken up by the two neighboring cells in row $N-1$ and another two are taken up by edge cells in row $N$. 

Now consider the unshared vertices. Every unshared vertex in row $N-1$ contributes $k-3$ vertex cells to row $N$, since one cell around the vertex is taken up by the cell in row $N-1$ and another two are taken up by edge cells in row $N$. This must be multiplied by the number of unshared vertices, which depends on the type of cell. Edge cells in row $N-1$ have $n-4$ unshared vertices, since two vertices belong to the outside of row $N-2$ and another two are shared with neighboring cells in row $N-1$. Thus unshared vertices from edge cells in row $N-1$ contribute $(n-4)(k-3)$ new vertex cells to row $N$. Vertex cells in row $N-1$ have $n-3$ unshared vertices, since one vertex belongs to the outside of row $N-2$ and another two are shared with neighboring cells in row $N-1$. Vertex cells in row $N-1$ then contribute $(n-3)(k-3)$ new vertex cells to row $N$.

The contributions from shared and unshared vertices are then added to get the total contributions of new vertex cells. Edge cells in row $N-1$ then contribute $(n-4)(k-3) + (k-4)$ new vertex cells to row $N$, while vertex cells in row $N-1$ contribute $(n-3)(k-3) + (k-4)$ to new vertex cells in row $N$. This fully determines the bottom row of $M$, completing our definition of $M$,
\begin{equation}
    M = 
    \begin{pmatrix}
        n-3                 &   n-2 \\
        (n-4)(k-3) + (k-4)  &   (n-3)(k-3) + (k-4)
    \end{pmatrix}
\end{equation}
and reproducing equation (\ref{eq:facevec}) from Sec.~\ref{sec:tessellations}.

\bibliographystyle{utphys}
\bibliography{Interior_TNs}

\end{document}

%% file: colors.tex
\scalebox{1.2}{
\begin{tikzpicture}[line width = 1.1pt]

\draw[purple] (0,3) -- (4,3);
\node (A) at (5,3) {$B$};
\draw[blue] (0,2.5) -- (4,2.5);
\node[below=0.5cm of A.west,anchor=west] (B) {$R_\text{out}$};
\draw[green] (0,2) -- (4,2);
\node[below=0.5cm of B.west,anchor=west] (C) {$r$};
\draw[orange] (0,1.5) -- (4,1.5);
\node[below=0.5cm of C.west,anchor=west] (D) {$\ell$};
\draw[red] (0,1) -- (4,1);
\node[below=0.5cm of D.west,anchor=west] (E) {$R_\text{in}$};

\end{tikzpicture}
}

%% file: fun_dyn.tex

\scalebox{1.15}{
\begin{tikzpicture}[line width=1.1pt]

\draw[red] (0,0) -- (0,0.5);
\draw[red] (0.5,0) -- (0.5,0.5);
\draw[red] (1,0) -- (1,0.5);
\draw[red] (1.5,0) -- (1.5,0.5);
\draw (2,0) -- (2,0.5);
\draw (2.5,0) -- (2.5,0.5);

\node at (0.75,-0.35) {$R_\text{in}$};
\node at (2.25,-0.35) {$|\psi\rangle_f$};

\draw[red] (0,0.5) -- (0,1.5);
\draw[red] (0.5,0.5) -- (0.5,1.5);
\draw[purple] (1,1) -- (1,1.5);
\draw[purple] (1.5,1) -- (1.5,1.5);
\draw[purple] (2,1) -- (2,1.5);
\draw[purple] (2.5,1) -- (2.5,1.5);

\draw[fill=gray!30] (0.75,0.5) rectangle (2.75,1);
\node at (1.75,0.75) {$U_0$};
\node at (-1,0.75) {$t=0$};

\draw[red] (0,1.5) -- (0,2.5);
\draw[purple] (0.5,2) -- (0.5,2.5);
\draw[purple] (1,2) -- (1,2.5);
\draw[purple] (1.5,2) -- (1.5,2.5);
\draw[blue] (2,2) -- (2,2.5);
\draw[blue] (2.5,2) -- (2.5,2.5);

\draw[fill=gray!30] (0.25,1.5) rectangle (2.75,2);
\node at (1.5,1.75) {$U_1$};
\node at (-1,1.75) {$t=1$};

\draw[purple] (0,3) -- (0,3.5);
\draw[purple] (0.5,3) -- (0.5,3.5);
\draw[blue] (1,3) -- (1,3.5);
\draw[blue] (1.5,3) -- (1.5,3.5);
\draw[blue] (2,2.5) -- (2,3.5);
\draw[blue] (2.5,2.5) -- (2.5,3.5);

\draw[fill=gray!30] (-0.25,2.5) rectangle (1.75,3);
\node at (0.75,2.75) {$U_2$};
\node at (-1,2.75) {$t=2$};

\node at (0.25,3.85) {$B$};
\node at (1.75,3.85) {$R_\text{out}$};

\end{tikzpicture}
}

%% file: eff_dyn.tex

\scalebox{1.2}{
\begin{tikzpicture}[line width=1.1pt]

\draw[red] (0,0) -- (0,0.5);
\draw[red] (0.5,0) -- (0.5,0.5);
\draw[red] (1,0) -- (1,0.5);
\draw[red] (1.5,0) -- (1.5,0.5);

\node at (0.75,-0.35) {$R_\text{in}$};

\draw[red] (0,0.5) -- (0,1.5);
\draw[red] (0.5,0.5) -- (0.5,1.5);
\draw[orange] (1,1) -- (1,1.5);
\draw[orange] (1.5,1) -- (1.5,1.5);

\draw[brown,fill=brown!20] (0.75,0.5) rectangle (1.75,1);
\node[brown] at (1.25,0.75) {$\hat{U}_0$};
\node at (-1,0.75) {$t=0$};

\draw[red] (0,1.5) -- (0,2.5);
\draw[orange] (0.5,2) -- (0.5,2.5);
\draw[orange] (1,2) -- (1,2.5);
\draw[orange] (1.5,2) -- (1.5,2.5);
\draw[green] (3.75,1.5) -- (2,2) -- (2,2.5);
\draw[green] (3.75,1.643) -- (2.5,2) -- (2.5,2.5);
\draw[blue] (3.75,1.5) -- (5.5,2) -- (5.5,2.5);
\draw[blue] (3.75,1.643) -- (5,2) -- (5,2.5);

\node at (3.75,1.15) {$|\text{MAX}\rangle_{r,R_\text{out}}$};
\draw[brown, fill=brown!20] (0.25,1.5) rectangle (1.75,2);
\node[brown] at (1,1.75) {$\hat{U}_1$};
\node at (-1,1.75) {$t=1$};

\draw[orange] (0,3) -- (0,3.5);
\draw[orange] (0.5,3) -- (0.5,3.5);
\draw[orange] (1,3) -- (1,3.5);
\draw[orange] (1.5,3) -- (1.5,3.5);
\draw[green] (2,3) -- (2,3.5);
\draw[green] (2.5,3) -- (2.5,3.5);
\draw[green] (3.75,2.5) -- (3,3) -- (3,3.5);
\draw[green] (3.75,2.833) -- (3.5,3) -- (3.5,3.5);
\draw[blue] (3.75,2.5) -- (4.5,3) -- (4.5,3.5);
\draw[blue] (3.75, 2.833) -- (4,3) -- (4,3.5);
\draw[blue] (5,2.5) -- (5,3.5);
\draw[blue] (5.5,2.5) -- (5.5,3.5);

\draw[brown, fill=brown!20] (-0.25,2.5) rectangle (2.75,3);
\node[brown] at (1.25,2.75) {$\hat{U}_2$};
\node at (-1,2.75) {$t=2$};

\node at (0.75,3.85) {$\ell$};
\node at (2.75,3.85) {$r$};
\node at (4.75,3.85) {$R_\text{out}$};

\end{tikzpicture}
}

%% file: transformations.tex

\scalebox{0.8}{
\begin{tikzpicture}[line width=1.1pt]

\node[scale=1.5] at (-1.3,2) {$V_\text{BF} =$};

\draw[orange] (0,-0.25) -- (0,0.5);
\draw[green] (2,-0.25) -- (2,0.5);
\draw[blue] (3,-0.25) -- (3,2) -- (2.5,2.25);

\draw[brown,fill=brown!20] (-0.25,0.5) rectangle (2.25,1.25);
\node[brown] at (1,0.875) {$\hat{U}^\dagger$};

\draw[red] (0,1.25) -- (0,2.75);
\draw[green] (2,1.25) -- (2,2) -- (2.5,2.25);
\node[font=\small] at (2.5,2.5) {$\langle\text{MAX}|_{r,R_\text{out}}$};
\draw (1,2.25) -- (1,2.75);
\node[font=\small] at (1,1.95) {$|\psi\rangle_f$};

\draw[fill=gray!30] (-0.25,2.75) rectangle (1.25,3.5);
\node at (0.5,3.125) {$U$};

\draw[purple] (0,3.5) -- (0,4.25);
\draw[blue] (1,3.5) -- (1,4.25);

\draw[->, line width=1.1mm] (4.25,2) -- (5.25,2);


\begin{scope}[shift={(6.5,0)}]

\draw[orange] (0,-0.25) -- (0,0.5);
\draw[green] (2,-0.25) -- (2,0.5);
\draw[blue] (3,-0.25) -- (3,2) -- (2.5,2.25);

\draw[brown,fill=brown!20] (-0.25,0.5) rectangle (2.25,1.25);
\node[brown] at (1,0.875) {$\hat{U}^\dagger$};

\draw[red] (0,1.25) -- (0,2.75);
\draw[green] (2,1.25) -- (2,2) -- (2.5,2.25);
\node[font=\small] at (2.5,2.5) {$\langle\text{MAX}|_{r,R_\text{out}}$};
\draw (1,2.25) -- (1,2.75);
\node[font=\small] at (1,1.95) {$|\psi\rangle_f$};

\draw[fill=gray!30] (-0.25,2.75) rectangle (1.25,3.5);
\node at (0.5,3.125) {$U$};

\draw[purple] (0,3.5) -- (0,4.25);
\draw[blue] (1,3.5) -- (1,4.25) -- (1.5,4.5);
\draw[green] (1.5,4.5) -- (2,4.25) -- (2,3.5) -- (2.5,3.25);
\draw[blue] (2.5,3.25) -- (3,3.5) -- (3,4.25);

\node[font=\small] at (1.5,4.75) {$|r|\langle\text{MAX}|_{r,R'_\text{out}}$};
\node[font=\small] at (2.5,3) {$|\text{MAX}\rangle_{r,R_\text{out}}$};

\end{scope}

\draw[->, line width=1.1mm] (10.75,2) -- (11.75,2);


\begin{scope}[shift={(13,0)}]

\draw[orange] (0,-0.25) -- (0,0.5);
\draw[green] (2,-0.25) -- (2,0.5);
\draw[blue] (3,-0.25) -- (3,4.25);

\draw[brown,fill=brown!20] (-0.25,0.5) rectangle (2.25,1.25);
\node[brown] at (1,0.875) {$\hat{U}^\dagger$};

\draw[red] (0,1.25) -- (0,2.75);
\draw[green] (2,1.25) -- (2,4.25) -- (1.5,4.5);
\draw (1,2.25) -- (1,2.75);
\node[font=\small] at (1,1.95) {$|\psi\rangle_f$};

\draw[fill=gray!30] (-0.25,2.75) rectangle (1.25,3.5);
\node at (0.5,3.125) {$U$};

\draw[purple] (0,3.5) -- (0,4.25);
\draw[blue] (1,3.5) -- (1,4.25) -- (1.5,4.5);
\node[font=\small] at (1.5,4.75) {$|r|\langle\text{MAX}|_{r,R'_\text{out}}$};

\node[scale=1.5] at (4.3,2) {$= V_\text{BFP}$};
    
\end{scope}

\end{tikzpicture}
}